\input harvmac


\input epsf

\newcount\figno
\figno=0
\def\fig#1#2#3{
\par\begingroup\parindent=0pt\leftskip=1cm\rightskip=1cm\parindent=0pt
\baselineskip=11pt
\global\advance\figno by 1
\midinsert
\epsfxsize=#3
\centerline{\epsfbox{#2}}
\vskip 12pt
{\bf Figure \the\figno:} #1\par
\endinsert\endgroup\par
}
\def\figlabel#1{\xdef#1{\the\figno}}

\def\subsubsec#1{\bigskip\noindent{\it{#1}} \bigskip}
\def\del{\partial}

\def\doublefig#1#2#3#4#5{
\par\begingroup\parindent=0pt\leftskip=1cm\rightskip=1cm\parindent=0pt
\baselineskip=11pt
\global\advance\figno by 1
\midinsert
\epsfxsize=#4
\centerline{\epsfbox{#2}\hskip0.2in\epsfxsize=#5\epsfbox{#3}}
\vskip 12pt
{\bf Figure \the\figno:} #1\par
\endinsert\endgroup\par
}
\def\figlabel#1{\xdef#1{\the\figno}}


\noblackbox

\def\IZ{\relax\ifmmode\mathchoice
{\hbox{\cmss Z\kern-.4em Z}}{\hbox{\cmss Z\kern-.4em Z}}
{\lower.9pt\hbox{\cmsss Z\kern-.4em Z}} {\lower1.2pt\hbox{\cmsss
Z\kern-.4em Z}}\else{\cmss Z\kern-.4em Z}\fi}
\def\mod{{\rm mod}}

\font\cmss=cmss10 \font\cmsss=cmss10 at 7pt
\def\IR{\relax{\rm I\kern-.18em R}}
\def\inbar{\,\vrule height1.5ex width.4pt depth0pt}
\def\IC{\relax\hbox{$\inbar\kern-.3em{\rm C}$}}
\def\IR{\relax{\rm I\kern-.18em R}}
\def\IP{\relax{\rm I\kern-.18em P}}
\def\IZ{\relax{\rm Z\kern-.34em Z}}

\def\ie{{\it i.e.}}
\def\frac#1#2{{#1 \over #2}}


\def\journal#1&#2(#3){\unskip, \sl #1\ \bf #2 \rm(19#3) }
\def\andjournal#1&#2(#3){\sl #1~\bf #2 \rm (19#3) }

\def\d{\partial}

%

%
\catcode`\@=11
\def\slash#1{\mathord{\mathpalette\c@ncel{#1}}}
\overfullrule=0pt
\def\AA{{\cal A}}

\def\DD{{\cal D}}

\def\GG{{\cal G}}

\def\NN{{\cal N}}
\def\OO{{\cal O}}

\def\SS{{\cal S}}

\def\eps{\epsilon}

\def\underrel#1\over#2{\mathrel{\mathop{\kern\z@#1}\limits_{#2}}}

\catcode`\@=12

\def\({\left(}
\def\){\right)}
\def\[{\left[}
\def\]{\right]}
\def\<{\langle}
\def\>{\rangle}
\def\half{{1\over 2}}
\def\d{\partial}

\def\|{\biggl|}
\def\mod{{\rm mod}}

\def\bk{{\bf k}}
\def\bp{{\bf p}}
\def\bx{{\bf x}}

\def\bA{{\bf A}}
\def\by{{\bf y}}
\def\ooint{\relax{\int\kern-.9em ^{\rm O}\kern-.75em _{\rm O}}}
\def\uoint{\relax{\int\kern-.87em ^{\rm O}}}
\def\doint{\relax{\int\kern-.97em _{\rm O}}}

\def\tn#1#2{t^{[#1]}_{#2}}
\def\li2{{\rm Li_2}}

\def\ie{{\it i.e.}}
\def\eg{{\it e.g.}}


%


\lref\GrossKZ{
  D.~J.~Gross and P.~F.~Mende,
  ``The High-Energy Behavior of String Scattering Amplitudes,''
  Phys.\ Lett.\  B {\bf 197}, 129 (1987);
 ``String Theory Beyond the Planck Scale,''
  Nucl.\ Phys.\  B {\bf 303}, 407 (1988).
}

\lref\DrukkerZQ{
  N.~Drukker, D.~J.~Gross and H.~Ooguri,
 ``Wilson loops and minimal surfaces,''
  Phys.\ Rev.\  D {\bf 60}, 125006 (1999)
  [arXiv:hep-th/9904191].
}

\lref\AharonyXZ{
  O.~Aharony, A.~Fayyazuddin and J.~M.~Maldacena,
``The large N limit of N = 2,1 field theories from three-branes in
  F-theory,''
  JHEP {\bf 9807}, 013 (1998)
  [arXiv:hep-th/9806159].
}

\lref\KarchSH{
  A.~Karch and E.~Katz,
 ``Adding flavor to AdS/CFT,''
  JHEP {\bf 0206}, 043 (2002)
  [arXiv:hep-th/0205236].
}

\lref\KomargodskiER{
  Z.~Komargodski and S.~S.~Razamat,
  ``Planar quark scattering at strong coupling and universality,''
  arXiv:0707.4367 [hep-th].
}

\lref\AldayHR{
  L.~F.~Alday and J.~Maldacena,
  ``Gluon scattering amplitudes at strong coupling,''
  JHEP {\bf 0706}, 064 (2007)
  [arXiv:0705.0303 [hep-th]].
}

\lref\BuscherSK{
  T.~H.~Buscher,
  ``A Symmetry of the String Background Field Equations,''
  Phys.\ Lett.\  B {\bf 194}, 59 (1987).
  ``Path Integral Derivation of Quantum Duality in Nonlinear Sigma Models,''
  Phys.\ Lett.\  B {\bf 201}, 466 (1988).
  ``Quantum Corrections And Extended Supersymmetry In New Sigma Models,''
  Phys.\ Lett.\  B {\bf 159}, 127 (1985).
}

\lref\MironovQQ{
  A.~Mironov, A.~Morozov and T.~N.~Tomaras,
  ``On n-point Amplitudes in N=4 SYM,''
  arXiv:0708.1625 [hep-th].
}

\lref\GrossGE{
  D.~J.~Gross and J.~L.~Manes,
  ``The High-energy Behavior of Open String Scattering,''
  Nucl.\ Phys.\  B {\bf 326}, 73 (1989).
}

\lref\PolyakovCA{
  A.~M.~Polyakov,
  ``Gauge Fields As Rings Of Glue,''
  Nucl.\ Phys.\  B {\bf 164}, 171 (1980).
}

\lref\KorchemskySI{
  G.~P.~Korchemsky,
  ``Asymptotics of the Altarelli-Parisi-Lipatov Evolution Kernels of Parton
  Distributions,''
  Mod.\ Phys.\ Lett.\  A {\bf 4}, 1257 (1989).
}

\lref\KorchemskyXV{
  G.~P.~Korchemsky and G.~Marchesini,
  ``Structure function for large x and renormalization of Wilson loop,''
  Nucl.\ Phys.\  B {\bf 406}, 225 (1993)
  [arXiv:hep-ph/9210281].
}

\lref\KorchemskayaQP{
  I.~A.~Korchemskaya and G.~P.~Korchemsky,
  ``High-energy scattering in QCD and cross singularities of Wilson loops,''
  Nucl.\ Phys.\  B {\bf 437}, 127 (1995)
  [arXiv:hep-ph/9409446].
}

\lref\KotikovPM{
  A.~V.~Kotikov and L.~N.~Lipatov,
  ``NLO corrections to the BFKL equation in QCD and in supersymmetric gauge
  theories,''
  Nucl.\ Phys.\  B {\bf 582}, 19 (2000)
  [arXiv:hep-ph/0004008].
}

\lref\KruczenskiFB{
  M.~Kruczenski,
  ``A note on twist two operators in N = 4 SYM and Wilson loops in Minkowski
  signature,''
  JHEP {\bf 0212}, 024 (2002)
  [arXiv:hep-th/0210115].
}

\lref\GubserTV{
  S.~S.~Gubser, I.~R.~Klebanov and A.~M.~Polyakov,
  Nucl.\ Phys.\  B {\bf 636}, 99 (2002)
  [arXiv:hep-th/0204051].
  Y.~Makeenko,
  JHEP {\bf 0301}, 007 (2003)
  [arXiv:hep-th/0210256].
  S.~Frolov and A.~A.~Tseytlin,
  JHEP {\bf 0206}, 007 (2002)
  [arXiv:hep-th/0204226].
  S.~Frolov, A.~Tirziu and A.~A.~Tseytlin,
  Nucl.\ Phys.\  B {\bf 766}, 232 (2007)
  [arXiv:hep-th/0611269].
  R.~Roiban, A.~Tirziu and A.~A.~Tseytlin,
  JHEP {\bf 0707}, 056 (2007)
  [arXiv:0704.3638 [hep-th]].
}

\lref\BernZX{
  Z.~Bern, L.~J.~Dixon, D.~C.~Dunbar and D.~A.~Kosower,
  ``One loop n point gauge theory amplitudes, unitarity and collinear limits,''
  Nucl.\ Phys.\  B {\bf 425}, 217 (1994)
  [arXiv:hep-ph/9403226].
}

\lref\BrandhuberYX{
  A.~Brandhuber, P.~Heslop and G.~Travaglini,
  ``MHV Amplitudes in N=4 Super Yang-Mills and Wilson Loops,''
  arXiv:0707.1153 [hep-th].
}

\lref\MagneaZB{
  L.~Magnea and G.~Sterman,
  ``Analytic continuation of the Sudakov form-factor in QCD,''
  Phys.\ Rev.\  D {\bf 42}, 4222 (1990).
}

\lref\dixonreviews{
 \eg\  Z.~Bern, L.~J.~Dixon and D.~A.~Kosower,
  ``Progress in one-loop QCD computations,''
  Ann.\ Rev.\ Nucl.\ Part.\ Sci.\  {\bf 46}, 109 (1996)
  [arXiv:hep-ph/9602280];
  L.~J.~Dixon,
  ``Calculating scattering amplitudes efficiently,''
  arXiv:hep-ph/9601359,
  and references therein and thereto.
  }

\lref\BernKQ{
  Z.~Bern, L.~J.~Dixon and D.~A.~Kosower,
  ``N = 4 super-Yang-Mills theory, QCD and collider physics,''
  Comptes Rendus Physique {\bf 5}, 955 (2004)
  [arXiv:hep-th/0410021].
}

\lref\CollinsGX{
  J.~C.~Collins, D.~E.~Soper and G.~Sterman,
  ``Factorization of Hard Processes in QCD,''
  Adv.\ Ser.\ Direct.\ High Energy Phys.\  {\bf 5}, 1 (1988)
  [arXiv:hep-ph/0409313].
}

\lref\sudakovreferences{ V. Sudakov, Sov. Phys. JETP {\bf 3} , 65
(1956). R. Jackiw, Ann. Phys. (N.Y.) {\bf 48}, 292 (1968).
   A.~H.~Mueller,
  Phys.\ Rev.\  D {\bf 20}, 2037 (1979).
   J.~C.~Collins,
  Phys.\ Rev.\  D {\bf 22}, 1478 (1980).
 A.~Sen,
  Phys.\ Rev.\  D {\bf 24}, 3281 (1981).
  }

\lref\sudakovreview{

  J.~C.~Collins,
  ``Sudakov form factors,''
  Adv.\ Ser.\ Direct.\ High Energy Phys.\  {\bf 5}, 573 (1989)
  [arXiv:hep-ph/0312336]; L.~Magnea and G.~Sterman,
  ``Analytic continuation of the Sudakov form-factor in QCD,''
  Phys.\ Rev.\  D {\bf 42}, 4222 (1990).
}

\lref\Korchsudakov{ G.~P.~Korchemsky and A.~V.~Radyushkin,
  Phys.\ Lett.\  B {\bf 171}, 459 (1986);
G. P. Korchemsky, and A. V. Radyushkin, Nucl. Phys. {\bf B283}, 342
(1987);
  G.~P.~Korchemsky,
  Phys.\ Lett.\  B {\bf 220}, 629 (1989).
  S.~V.~Ivanov, G.~P.~Korchemsky and A.~V.~Radyushkin,
  Yad.\ Fiz.\  {\bf 44}, 230 (1986)
  [Sov.\ J.\ Nucl.\ Phys.\  {\bf 44}, 145 (1986)].
}

\lref\catani{
  S.~Catani,
  ``The singular behaviour of {QCD} amplitudes at two-loop order,''
  Phys.\ Lett.\  B {\bf 427}, 161 (1998)
  [arXiv:hep-ph/9802439]; G.~Sterman and M.~E.~Tejeda-Yeomans,
  ``Multi-loop amplitudes and resummation,''
  Phys.\ Lett.\  B {\bf 552}, 48 (2003)
  [arXiv:hep-ph/0210130].
}

\lref\AldayMF{
  L.~F.~Alday and J.~Maldacena,
 ``Comments on operators with large spin,''
  arXiv:0708.0672 [hep-th].
}

\lref\sceft{ C.~W.~Bauer, S.~Fleming and M.~E.~Luke,
  Phys.\ Rev.\  D {\bf 63}, 014006 (2001)
  [arXiv:hep-ph/0005275];
  C.~W.~Bauer, S.~Fleming, D.~Pirjol and I.~W.~Stewart,
  Phys.\ Rev.\  D {\bf 63}, 114020 (2001)
  [arXiv:hep-ph/0011336];
  C.~W.~Bauer, D.~Pirjol and I.~W.~Stewart,
  Phys.\ Rev.\ Lett.\  {\bf 87}, 201806 (2001)
  [arXiv:hep-ph/0107002];
 C.~W.~Bauer, D.~Pirjol and I.~W.~Stewart,
  Phys.\ Rev.\  D {\bf 65}, 054022 (2002)
  [arXiv:hep-ph/0109045];
  A.~V.~Manohar,
  Phys.\ Rev.\  D {\bf 68}, 114019 (2003)
  [arXiv:hep-ph/0309176].
}

\lref\bds{
  Z.~Bern, L.~J.~Dixon and V.~A.~Smirnov,
  ``Iteration of planar amplitudes in maximally supersymmetric Yang-Mills
  theory at three loops and beyond,''
  Phys.\ Rev.\  D {\bf 72}, 085001 (2005)
  [arXiv:hep-th/0505205].
}


\lref\MS{
  J.~McGreevy, and A.~Sever,
in progress.
}

\lref\SchnitzerRN{
  H.~J.~Schnitzer,
  ``Reggeization of N=8 Supergravity and N=4 Yang-Mills Theory II,''
  arXiv:0706.0917 [hep-th].
}

\lref\BurringtonID{
  B.~A.~Burrington, J.~T.~Liu, L.~A.~Pando Zayas and D.~Vaman,
  ``Holographic duals of flavored N = 1 super Yang-Mills: Beyond the probe
  approximation,''
  JHEP {\bf 0502}, 022 (2005)
  [arXiv:hep-th/0406207].
}

\lref\KruczenskiBE{
  M.~Kruczenski, D.~Mateos, R.~C.~Myers and D.~J.~Winters,
  ``Meson spectroscopy in AdS/CFT with flavour,''
  JHEP {\bf 0307}, 049 (2003)
  [arXiv:hep-th/0304032].
}

\lref\BernFZ{
  Z.~Bern, L.~J.~Dixon and D.~A.~Kosower,
  ``One Loop Corrections To Two Quark Three Gluon Amplitudes,''
  Nucl.\ Phys.\  B {\bf 437}, 259 (1995)
  [arXiv:hep-ph/9409393].
}

\lref\BernCT{
  Z.~Bern, J.~J.~M.~Carrasco, H.~Johansson and D.~A.~Kosower,
  ``Maximally supersymmetric planar Yang-Mills amplitudes at five loops,''
  arXiv:0705.1864 [hep-th].
}

\lref\otherrefs{

    S.~Abel, S.~Forste and V.~V.~Khoze,
  arXiv:0705.2113 [hep-th];
  E.~I.~Buchbinder,
  arXiv:0706.2015 [hep-th];
  R.~C.~Brower, M.~J.~Strassler and C.~I.~Tan,
  arXiv:0707.2408 [hep-th];
    M.~Kruczenski, R.~Roiban, A.~Tirziu and A.~A.~Tseytlin,
  arXiv:0707.4254 [hep-th];
     A.~Jevicki, C.~Kalousios, M.~Spradlin and A.~Volovich,
  arXiv:0708.0818 [hep-th];
      D.~Nguyen, M.~B.~Spradlin and A.~Volovich,
  arXiv:0709.4665 [hep-th].
}

\lref\DrummondCF{
  J.~M.~Drummond, J.~Henn, G.~P.~Korchemsky and E.~Sokatchev,
  ``On planar gluon amplitudes/Wilson loops duality,''
  arXiv:0709.2368 [hep-th].
}

\lref\BeisertEZ{
  N.~Beisert, B.~Eden and M.~Staudacher,
  J.\ Stat.\ Mech.\  {\bf 0701}, P021 (2007)
  [arXiv:hep-th/0610251];
  M.~K.~Benna, S.~Benvenuti, I.~R.~Klebanov and A.~Scardicchio,
  Phys.\ Rev.\ Lett.\  {\bf 98}, 131603 (2007)
  [arXiv:hep-th/0611135];
  A.~V.~Kotikov and L.~N.~Lipatov,
  Nucl.\ Phys.\  B {\bf 769}, 217 (2007)
  [arXiv:hep-th/0611204];
  L.~F.~Alday, G.~Arutyunov, M.~K.~Benna, B.~Eden and I.~R.~Klebanov,
  JHEP {\bf 0704}, 082 (2007)
  [arXiv:hep-th/0702028];
  I.~Kostov, D.~Serban and D.~Volin,
  [arXiv:hep-th/0703031]; P.~Y.~Casteill and C.~Kristjansen,
  Nucl.\ Phys.\  B {\bf 785}, 1 (2007)
  [arXiv:0705.0890 [hep-th]]; B.~Basso, G.~P.~Korchemsky and J.~Kotanski,
  arXiv:0708.3933 [hep-th].
}

\lref\Mandelstam{
 S.\ Mandelstam, Phys.\ Rev.\ {\bf B137} (1965)
949; H.~J.~Schnitzer, ``Reggeization of N = 8 supergravity and N = 4
Yang-Mills theory,'' arXiv:hep-th/0701217; M.T.\ Grisaru, H.J.\
Schnitzer and H-S.\ Tsao, ``The Reggeization of Yang--Mills gauge
mesons in theories with a spontaneously broken symmetry," Phys.\
Rev.\ Lett.\ {\bf 20} (1973) 811; ``Reggeization of elementary
particles in renormalizable gauge theories:  vectors and spinors,"
Phys.\ Rev.\  {\bf D8} (1973) 4498; ``The Reggeization of elementary
particles in renormalizable gauge theories:  scalars," Phys.\ Rev.\
{\bf D9} (1974) 2864; M.T.\ Grisaru and H.J.\ Schnitzer,
``Reggeization of gauge vector mesons and unified theories," Phys.\
Rev.\ {\bf D20} (1979) 784; ``Reggeization of elementary fermions in
arbitrary renormalizable gauge theories," Phys.\ Rev.\ {\bf D21}
(1980) 1952.}

\lref\juanstrings{
J.~Maldacena, talk at Strings 2007.
}

\lref\NaculichUB{
  S.~G.~Naculich and H.~J.~Schnitzer,
  ``Regge behavior of gluon scattering amplitudes in N=4 SYM theory,''
  arXiv:0708.3069 [hep-th].
}

\lref\DrummondAU{
  J.~M.~Drummond, G.~P.~Korchemsky and E.~Sokatchev,
  ``Conformal properties of four-gluon planar amplitudes and Wilson loops,''
  arXiv:0707.0243 [hep-th].
}

\lref\StermanWJ{
  G.~Sterman and S.~Weinberg,
  ``Jets From Quantum Chromodynamics,''
  Phys.\ Rev.\ Lett.\  {\bf 39}, 1436 (1977).
}

\lref\AldayHE{
  L.~F.~Alday and J.~M.~Maldacena,
  arXiv:0710.1060 [hep-th].
}

\Title{\vbox{\baselineskip12pt
\hbox{arXiv:0710.0393}\hbox{BRX TH-591}\hbox{MIT-CTP/3871}
}} {\vbox{ \centerline{Quark scattering amplitudes} \centerline{at
strong coupling}
\smallskip
\smallskip
\smallskip
}} \centerline{John McGreevy${}^1$ and Amit Sever${}^2$}
\bigskip
\centerline{{$^1$Center for Theoretical Physics, Massachussetts
Institute of Technology}} \centerline{{Cambridge, MA 02139, USA}}
\centerline{{$^2$Brandeis Theory Group, Martin Fisher School of
Physics, Brandeis University}} \centerline{{Waltham, MA 02454, USA}}

\bigskip
\bigskip
\bigskip
\noindent

Following Alday and Maldacena \AldayHR, we describe a string theory
method to compute the strong coupling behavior of the scattering
amplitudes of quarks and gluons in planar $\NN=2$ super Yang-Mills
theory in the probe approximation. Explicit predictions for these
quantities can be constructed using the all-orders planar gluon
scattering amplitudes of $\NN=4$ super Yang-Mills due to Bern, Dixon
and Smirnov \bds.

\Date{October, 2007}

\newsec{Introduction}

A great deal of effort has been devoted to the
computation of scattering amplitudes
of massless partons in non-abelian gauge theories \dixonreviews.
Such amplitudes are IR divergent and require
a regulator,
which is always present in a physical observable
\refs{\eg\ \StermanWJ}.
Via factorization theorems \CollinsGX, these
quantities are used to make predictions for high-energy QCD
collision events. Amplitudes with both gluons and quarks are
required for precise theoretical predications at high-energy
colliders.

Some of the progress in constructing such amplitudes is facilitated
by supersymmetry \refs{\dixonreviews, \BernKQ}. $S$-matrix elements
for massless partons in supersymmetric gauge theories are useful in
a number of ways for physical high-energy QCD calculations. Firstly,
perturbative SYM scattering amplitudes share many qualitative
properties with QCD amplitudes in the regime relevant for jet
physics. The Supersymmetric Yang-Mills (SYM) answers, however, are
simpler than the QCD amplitudes, and hence are useful in learning to
understand their structure. More pragmatically, tree level
amplitudes with quarks can be related to supersymmetric amplitudes
with gluinos using color manipulations; supersymmetry Ward
identities then relate them to gluon amplitudes. SYM theories can be
used as a testing ground for computational techniques. Finally,
amplitudes in supersymmetric theories are useful as building blocks
for those of QCD \BernKQ.

Work on planar gluon scattering in the $\NN=4$ theory has culminated
in a conjecture by Bern, Dixon and Smirnov (BDS) \bds\ for the
all-orders $n$-point planar MHV amplitudes. Very recently, Alday and
Maldacena (AM) uncovered some of this structure in the string theory
dual description \AldayHR. In their ground-breaking paper, the
authors show that the gauge theory S-matrix can be defined and
regulated from the AdS string theory. The S-matrix is controlled at
large 't Hooft coupling $\lambda$ by a classical worldsheet. AM
match the Sudakov IR divergence structure using local features of
the extremal worldsheet. They also find the explicit worldsheet
configuration for four gluons, using a solution of \KruczenskiFB.
When combined with knowledge of the strong-coupling behavior of the
cusp anomalous dimension \refs{\GubserTV,\KruczenskiFB,\BeisertEZ},
the area of the four-gluon worldsheet can be seen to match the
strong coupling momentum-dependence predicted by the BDS ansatz.

Finding other explicit worldsheet solutions seems to be quite
difficult. In order to understand the purview and utility of the AM
procedure, it is worthwhile to try to extend it to other theories.
The simplicity of the strong-coupling limit implied by the existence
of a perturbative string theory description extends also to models
with matter in the fundamental representation, such as those of
\refs{\AharonyXZ\KarchSH}. While tree level amplitudes with quarks can be
reconstructed from gluon amplitudes as described above, this is no
longer true at loop level, and new types of behavior can emerge. It
would therefore be valuable to extend the strong-coupling
description to such theories.

In this paper, we generalize the prescription of \AldayHR\ to the
holographic dual of a gauge theory with matter in the fundamental
representation of the gauge group, which we will generically call quarks.
The model we study arises by adding a small number of D7-branes to
$AdS_5\times S^5$ in a way that preserves eight supercharges
\refs{\AharonyXZ,\KarchSH}. At strong coupling, scattering amplitudes in the theory
with flavor are governed a simple generalization of the
minimal-surface problem studied by AM.

Our most interesting result is a relation between field theories.
From a symmetry relation between their strong-coupling dual
worldsheets, we will find a set of equations relating gluon
scattering in the $\NN=4$ theory and quark-and-gluon scattering in
the flavored theory. We can then exploit the conjecture of BDS \bds\
to make predictions for quark scattering amplitudes.

The paper is organized as follows. In section two and appendix A, we
review the conjecture of BDS for planar gluon scattering. In section
three, we give a detailed exposition of the prescription of Alday
and Maldacena. In the course of reviewing the method, we explain how
the positions of the vertex operators are determined by the T-dual
solution\foot{We thank Hong Liu for asking this question.}. In
section four, we explain the generalization of the prescription to
include fields in the fundamental representation of the gauge group.
Section five describes a relationship between quark and gluon
scattering which follows from their string theory descriptions.
Using this relationship and the BDS ansatz, we make an explicit
prediction for the $\bar qggq$ amplitude. In section six, we study
the IR divergences of the resulting quark scattering amplitudes, and
verify the divergent part of the result with a direct argument in
the gauge theory with fundamentals. In section seven we discuss the
comparison of our prediction with perturbative $\NN=2$ results.
Toward that aim, we also study the Regge limit of quark scattering
amplitudes in $\NN=2$ SYM.

\newsec{Gluon scattering in planar $\NN=4$ SYM}

Based on explicit computation of the planar four-gluon scattering
amplitude up to four loops, Bern, Dixon and Smirnov (BDS) gave an
ansatz for the all-loop planar n-gluon amplitudes $\AA_n$ in $\NN=4$
SYM (at least, for the maximally helicity violating ones) \bds. At
strong coupling, Alday and Maldacena (AM) have confirmed the BDS
ansatz for the 4-gluon amplitude (for the $in\to out\to in\to out$
momentum ordering) using the AdS/CFT correspondence \AldayHR. The
form of the BDS ansatz for the four gluon amplitude can be described
graphically as in Fig.\ 1.
\vskip-.1in \fig{The BDS ansatz for the four gluon amplitude, where
$f(\lambda)$ is some function of the coupling.}{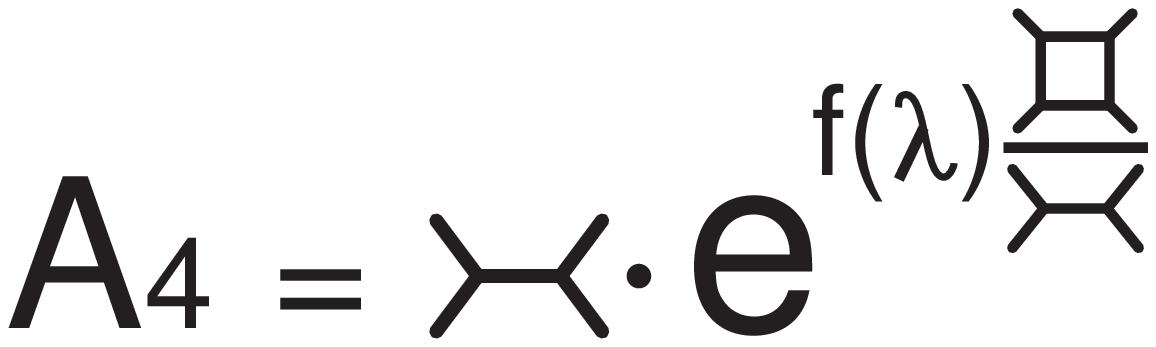}{1in}
More precisely, the gluon amplitude is IR divergent and needs an
infrared regulator. In dimensional regularization
$d=4-2\epsilon$,\foot{More precisely, in {\it
four-dimensional-helicity} scheme, in which all helicity states are
four-dimensional and only the loop momentum is continued to
$d=4-2\epsilon$ dimensions.} the BDS ansatz for the n-gluon
amplitude is
\eqn\ansatz{\AA_n=\AA_n^{\rm tree}~ e^{-\SS_n}~,}
where $\AA_n^{\rm tree}$ is the tree-level amplitude and
\eqn\Sn{-\SS_n=\sum_{l=1}^{\infty}\lambda_\epsilon^l\(f^{(l)}(\epsilon)
M_{n}^{(1)}(l \epsilon) + C^{(l)}+E_n^{(l)}(\epsilon )\)}
does not depend on any color or helicity factors.
The symbols appearing in \Sn\ are defined as follows.
$M_n^{(1)}=\AA_n^{(1)}/\AA_n^{\rm tree}$ is the
ratio of one-loop and tree amplitudes.
\eqn\thooft{\lambda_\epsilon={g^2
N\over8\pi^2}\({4\pi\over\mu^2e^\gamma}\)^\eps~,\qquad\gamma=-\Gamma'(1)}
is the 't Hooft coupling\foot{Note that for $\epsilon\ne0$ the gauge
coupling $g$ is dimensionful, whereas $\lambda_\epsilon$ is
dimensionless.}, and $\mu$ is an IR cutoff.
 $f^{(l)}(\epsilon ) =
f_0^{(l)} + f_1^{(l)} \epsilon + f_2^{(l)} \epsilon^2$ is a set of
functions, one at each loop order, which make their appearance in
the exponentiated all-loop expression for the infrared divergences
in generic amplitudes 
\MagneaZB. In particular, $f(\lambda)\equiv 4
\sum_lf^{(l)}_0\lambda^l$ is the cusp anomalous dimension (equal to
the anomalous dimension of twist-two operators of large spin).
Its large 't Hooft coupling asymptotic is $f(\lambda) \sim
\sqrt{\lambda} + \hbox{const} + O(1/\sqrt{\lambda})$
\refs{\GubserTV,\KruczenskiFB,\BeisertEZ}. An important aspect of
the conjecture is that the constants $C^{(l)}$ do not depend on
kinematics or on the number of particles $n$. The non-iterating
remainders $E_n^{(l)}$ vanish as $\epsilon \to 0$ and depend
explicitly on $n$.

$\SS_n$ can be decomposed into finite and divergent pieces as
$\SS_n=\SS_n^{div}+\SS_n^{fin}$, where
\eqn\div{-\SS_n^{div}=\sum_{l=1}^\infty
\lambda_\epsilon^lf^{(l)}(\epsilon)I_n^{(1)}(l\epsilon)}
with (the 1-loop scalar box integral)
\eqn\In{I_n^{(1)}(\epsilon)=-\half{1\over\epsilon^2}\sum_{i=1}^n\({\mu^2\over
-s_{i,i+1}}\)^{\epsilon} .}
$s_{i,i+1}$ is {\it minus} the square of the sum of the $i$ and
$i+1$ external (outgoing) momenta, and as apposed to the convention
used in \bds, here we work with $(-+++)$ signature. $f_1^{(l)}\equiv
{l\over 2}\GG_0^{(l)}$ is the sub-leading divergent term. Expanding
\div\ in powers of $\epsilon$, in the limit $\epsilon\to 0$ it
gives\foot{The $1/\epsilon^2$ and $1/\epsilon$ terms cancel in
physical processes, as per the theorems of Bloch-Nordsieck and
Kinoshita-Lee-Nauenberg, \eg\ \StermanWJ.}
\eqn\divnoeps{-\SS_n^{div}=-{f(\lambda)\over 16}\ln^2\({\mu^2\over
-s_{i,i+1}}\)-{g(\lambda)\over 4}\ln\({\mu^2\over
-s_{i,i+1}}\)-\half h(\lambda)~,}
where $g(\lambda)=\sum\lambda^l \GG_0^{(l)}$ and
$h(\lambda)=\sum\lambda^lf^{(l)}_2/l^2$.\foot{Note that $g$ and $h$
change if we rescale the IR cutoff.}

The finite part can be written in terms of its 1-loop counterpart
$F^{(1)}_n$ as:
\eqn\finite{-\SS_n^{fin}={f(\lambda)\over 4} F^{(1)}_n+C(\lambda)~,}
where $C(\lambda)$ depends neither on $n$, nor on momenta. The
one-loop finite remainder, $F^{(1)}_n$ was evaluated in \BernZX.
Since in sections 5 and 7 we will need its explicit value, we write
it in Appendix A.

\subsubsec{Double-loop representation of the one-loop amplitude}

As was recently shown in
\refs{\DrummondAU, \BrandhuberYX}
the one-loop $n$-gluon
amplitude, $M_1^{(n)}$ (which contains both the divergent and finite
pieces), can be written as a simple double contour integral.
This expression
has a nice geometrical interpretation as the dimensionally
regularized, one-loop contribution to a polygonal Wilson loop $\Pi$,
made from the successive external momenta (see section 3):\foot{Note
the sign difference in the dimensional reduced Wilson loop
computation at $d=4-2\tilde\epsilon$, $\tilde\epsilon=-\epsilon>0$.
This can be understood as a consequence of the fact that T-duality
 interchanges the UV with the IR and
therefore the sign of $\epsilon$.} \foot{This surprising
relationship between field theory quantities is made somewhat less
shocking in light of the AM prescription, reviewed in the next
section. We record it here because it will be used in section 6.}
\foot{The relationship between the scattering
amplitude and the Wilson loop has recently been confirmed in more
detail in \DrummondCF.
}
\eqn\dloop{M^{(n)}_1 = \half\mu^{2\epsilon}\oint_\Pi\oint_\Pi
\frac{d\by\cdot d\by'} {[-(\by-\by')^2]^{1+\epsilon}}~.}
In \AldayHR, in addition to dimensional regularization, another
regularization that is more natural from the AdS/CFT point of view
was used. In AdS, it corresponds to cutting the Poincar\'e radial
coordinate at some small value $r_{IR}=1/z_{IR}$ and imposing the
boundary conditions there. The double loop integral \dloop, has a
natural adaptation to the AdS-regularization:
\eqn\dloopAdS{\widetilde M^{(n)}_1 = \half\oint_\Pi\oint_\Pi
\frac{d\by\cdot d\by'} {(\by-\by')^2+1/z_{IR}^2}~.}
However, we have not explicitly compared this expression
with the area of the AM 4-gluon
solution with a radial cutoff.
Note that although the boundaries of the polygon loop are
still null (and are therefore T-dual to null gluons), the particle
being exchanged is now massive.

In this paper we will {\it assume} that the BDS ansatz \ansatz\ is
correct and will combine it with its AdS dual picture to give a
prediction for the quark and gluon-quark planar amplitude at strong
coupling.

\newsec{The strong coupling dual of planar $\NN=4$ SYM gluon scattering}

In \AldayHR, Alday and Maldacena used the AdS/CFT correspondence to
compute gluon amplitudes in the t' Hooft limit of N = 4 SYM at
strong coupling.

On the string theory side, the leading order result at strong
coupling is given by a single classical string configuration
associated to the scattering process. As explained in \AldayHR, the
string theory scattering in AdS is happening at fixed angles and
large proper momentum and it is thus determined by a classical
solution. The final form for the color ordered planar scattering
amplitude of $n$ gluons at strong coupling is of the form
\eqn\stringA{\AA\sim e^{-S_{cl}} = e^{-{\sqrt\lambda\over 2\pi}({\rm
Area})_{cl}}~,}
where $S_{cl}$ denotes the action of a classical solution of the
string worldsheet equations of motion, which is proportional to the
area of the string world-sheet and
$\sqrt\lambda=R_{AdS}^2/l_s^2$.\foot{We set $\alpha'=1$.} The
solution depends on the 4-momenta, $\bk_i$, of the gluons. The whole
dependence on the coupling is in the overall factor.

In more detail, one has to compute the worldsheet path integral
over worldsheets with the topology of a disk, embedded in Poincar\'e
AdS:
\eqn\Poincare{ds^2=R_{AdS}^2{dz^2+dx_{3+1}^2\over z^2}~.}
The open strings have Neumann boundary conditions in the 3+1 spatial
directions and fixed radial position $z_{IR}$ at the worldsheet
boundary (\ie\ Dirichlet boundary condition in $z$). In addition one
should add open string insertions, dual to the asymptotic gluons
being scattered. These vertex operators have fixed dual field theory
4-momentum ${\bf k}$ along the spatial directions ${\bf x}$. At
$z_{IR}$, the proper string momentum (conjugate to the coordinates
$d\hat \bx=\lambda^{1/4}d\bx/z_{IR}$) is $\bk\,z_{IR}/\lambda^{1/4}$
and therefore, in a limit where $z_{IR}\to \infty$
faster than $\lambda^{1/4}$, the proper momentum diverges.
The problem becomes a classical problem of finding the saddle
point \GrossKZ\
of the worldsheet path integral.
We will discuss the validity of this saddle-point approximation
at the end of \S3.2.

\subsec{Using T-duality to simplify the boundary conditions}

It will be useful to review the solution of this steepest-descent
problem in some detail.
The open
string vertex operators are
\eqn\VO{V_{open}(\bk_i;\sigma_i)\propto  e^{i{\bk_i\cdot
\bx(\sigma_i)}}~,}
where $\sigma \in [-\infty, \infty)$
parameterizes the boundary of the disk
(upper half plane), $\bx(\sigma)=\bx(\sigma,\tau)|_{\tau=0}$ and
$\bk^2=0$. The vertex operators can be rewritten as contributions to
the boundary action as:
\eqn\ibp{\eqalign{&i\sum_{j=1}^n{\bf  k}_j\cdot{\bf
x}(\sigma_j)=i\sum_{j=1}^n{\bf k}_j\cdot\int d\sigma~{\bf
x}(\sigma)\delta(\sigma-\sigma_j)\cr =&i\sum_{j=1}^{n-1}{\bf
k}_j\cdot\int d\sigma{\bf
x}(\sigma)\d_\sigma\theta(\sigma;\sigma_j,\sigma_n)=-i\sum_{j=1}^{n-1}{\bf
k}_j\cdot\int d\sigma\d_\sigma{\bf
x}(\sigma)\[\theta(\sigma;\sigma_j,\sigma_n)+c\]\cr
=&-i\sum_{j=1}^n\int_{\sigma_j}^{\sigma_{j+1}}d\sigma\d_\sigma{\bf
x}(\sigma)\cdot\(\sum_{i\le j}\bk_i+{\bf c}\)~,}}
where $\theta$ is the periodic step function
\eqn\step{\theta(\sigma;\sigma_i,\sigma_j)=\left\{\matrix{1&
\sigma_i<\sigma<\sigma_j\cr 0&{\rm otherwise}}\right.~,}
$\sigma_1<\sigma_2<\dots<\sigma_n$, $\sigma_{n+1}=\sigma_1$ and
${\bf c}$ is a constant 4-vector.\foot{Note that since
$\sum_{j=1}^{n-1}{\bf k}_j=-{\bf k}_n$, the sum in the second line
of \ibp\ runs up to $n-1$ and we could equivalently choose to omit
any other ${\bf k}_{j<n}$ instead of ${\bf k}_n$.}

Next \AldayHR, as a technical trick to find the saddle point, we do
a change of variables in the path integral which can be described as
a ``T-duality" along the non-compact 3+1 flat directions. To do
this, we follow Buscher \BuscherSK. For each field $x^\mu$, we gauge
the shift symmetry $ x^\mu \to x^\mu + a$, and introduce a
worldsheet gauge field $A_\alpha^\mu$ and a scalar lagrange
multiplier $y^\mu$. We then consider the gauge-invariant action
\eqn\Bucher{S={\sqrt\lambda\over 4\pi}\int_\DD d\sigma\d\tau\
\[(\d_\alpha
\bx-\bA_\alpha)^2/z^2-i\by \cdot{\bf
F}\]+i\sum_{j=1}^{n-1}\int_{\sigma_j}^{\sigma_{j+1}}d\sigma
[\d_\sigma{\bf x}-\bA_\sigma]\cdot(\sum_{i\le j}\bk_i+{\bf c})~,}
where ${\bf F}=\d_\tau\bA_\sigma-\d_\sigma\bA_\tau$,
$\(\d_\tau\bx-\bA_\tau\)|_{\tau=0}=0$ and we are suppressing the
kinetic term for $z$. Now we can gauge fix $\bx=0$, so the action
becomes
\eqn\Buchertwo{S={\sqrt\lambda\over 4\pi}\int_\DD d\sigma\d\tau\
\[\bA_\alpha\cdot \bA_\alpha/z^2-i\by \cdot{\bf
F}\]-i\sum_{j=1}^{n-1}\int_{\sigma_j}^{\sigma_{j+1}}d\sigma {\bf
A}_\sigma\cdot(\sum_{i\le j}\bk_i+{\bf c})~.}
If we first integrate out $\by $, then we see that
$A_\alpha=-\d_\alpha\widetilde \bx$ is a flat connection and
therefore \Bucher\ is equivalent to the original action. If on the
other hand, we first integrate $\bA$, then it is convenient to
integrate by parts in the second term in \Buchertwo. We then
have
\eqn\Buchertwo{\eqalign{S=&{\sqrt\lambda\over 4\pi}\int_\DD
d\sigma\d\tau\
\[\bA_\alpha\cdot \bA_\alpha/z^2+i\(\bA_\sigma\cdot\d_\tau\by
-\bA_\tau\cdot\d_\sigma\by
\)\]\cr-&i\sum_{j=1}^n\int_{\sigma_j}^{\sigma_{j+1}}d\sigma {\bf
A}_\sigma\cdot(\sum_{i\le j}\bk_i+{\bf c}+{\sqrt\lambda\over
4\pi}\by )~.}}
It is convenient to rescale $\bA,\ \by$ and $z$ as
$\(\bA,\by,z\)\to\({\sqrt\lambda\over
4\pi}\bA,{4\pi\over\sqrt\lambda}\by,{\sqrt\lambda\over 4\pi} z\)$,
so that $\sqrt\lambda$ will stand in front of the whole action.
Integrating
out $\bA$ we find in the bulk of the disk an action for
($\by,\ r=1/z$) describing a dual AdS background\foot{Note that $r$
and $\by$ have dimensions of ${1\over \rm length}$.}~\foot{Note that
in terms of the original un-rescaled radial coordinate $z$, we have
$r=\sqrt\lambda/z$.}
\eqn\dualPoin{ds^2=R_{AdS}^2{dr^2+d\by\cdot d\by\over r^2}~.}

Integrating over the boundary value of $\bA_\sigma$ enforces the
boundary condition
\eqn\bc{\by(\sigma_i<\sigma<\sigma_{i+1})=-\sum_{j\le i}\bk_j-{\bf
c}~.}
The boundary condition \bc\ means that $\by$ is constant on any
segment $\sigma_i<\sigma<\sigma_{i+1}$ and jumps by $-{\bf k}_i$ at
$\sigma_i$. Due to the momentum conservation $\sum_i{\bf k}_i=0$,
the boundary of the string in the T-dual $\by$-coordinates is
closed. It stretches along a polygonal loop made of the ordered open
string momenta ${\bf k}_i$, centered at ${\bf c}$. ${\bf c}$ drops
out of the calculation and will be set to zero from now on.

In addition, the Gaussian integration over $\bA$ introduces a linear
dilaton ($\Phi\sim \log(z)$). It can be ignored in the classical
saddle point approximation because the dilaton term does not scale
as $\sqrt\lambda=R_{AdS}^2/l_s^2$; so as long as
$\sqrt\lambda\gg\log(z_{IR})$, it can be neglected with respect to
the kinetic and boundary terms.

Note that we have ``T-dualized" along non-compact directions as well
as time. That was possible because we were interested only in the
classical limit of the disk amplitude, as opposed to a quantum
mechanical object which would involve winding of closed strings.
The non-compact ``T-duality" may not be valid if one considers
worldsheets with non-trivial topology.

\subsec{The role of the vertex insertion points}

The solution to the equation of motion is an extremal-area surface
in the T-dual $\by$-coordinates subject to the boundary conditions
\bc. For any distribution of the ordered open string vertex
operators $\sigma_1<\sigma_2<\dots<\sigma_n$, there is such
a solution,
whose fluctuations are suppressed at large $\lambda$.
Physically, these extremal surfaces solution differ by the
momentum flow along the $\by$-segments which is controlled by the
values of the $\sigma$'s. After changing variables back to the $\bx$
space, this momentum flow translates into the extension (``winding")
of the asymptotic classical open string state. If the momenta
$\bk_i$ are null, then, as we will next prove, the solution with the
extremal surface area is the one which satisfies Neumann boundary
condition in the $\bk_i$ directions along the corresponding $\by$
segment. Therefore, for generic values of the $\sigma$'s, the
asymptotic open string states have non-zero extent (in the $\bx$
coordinates).\foot{See \MironovQQ\ for such explicit solutions to
the equations of motion, which end on the same polygonal locus but
have nonzero momentum flow through the boundary. \MironovQQ\ show
explicitly that the AM solution extremizes the area in this class of
solutions.} At the extremum they have zero extent (zero ``winding")
as one would physically expect (see Fig.\ 1).

\fig{A rough picture of the 4-gluon worldsheet embedding in $AdS_3$.
The worldsheet has a Dirichlet boundary condition in the radial
direction and may have asymptotic extent (``winding") in the other
directions, as indicated. The extremum of the path integral over the
vertex operator insertion points corresponds to the embedding where
the asymptotic string states have zero extent. It is strongly peaked
as the proper momentum $k z_{IR}\to\infty$.}{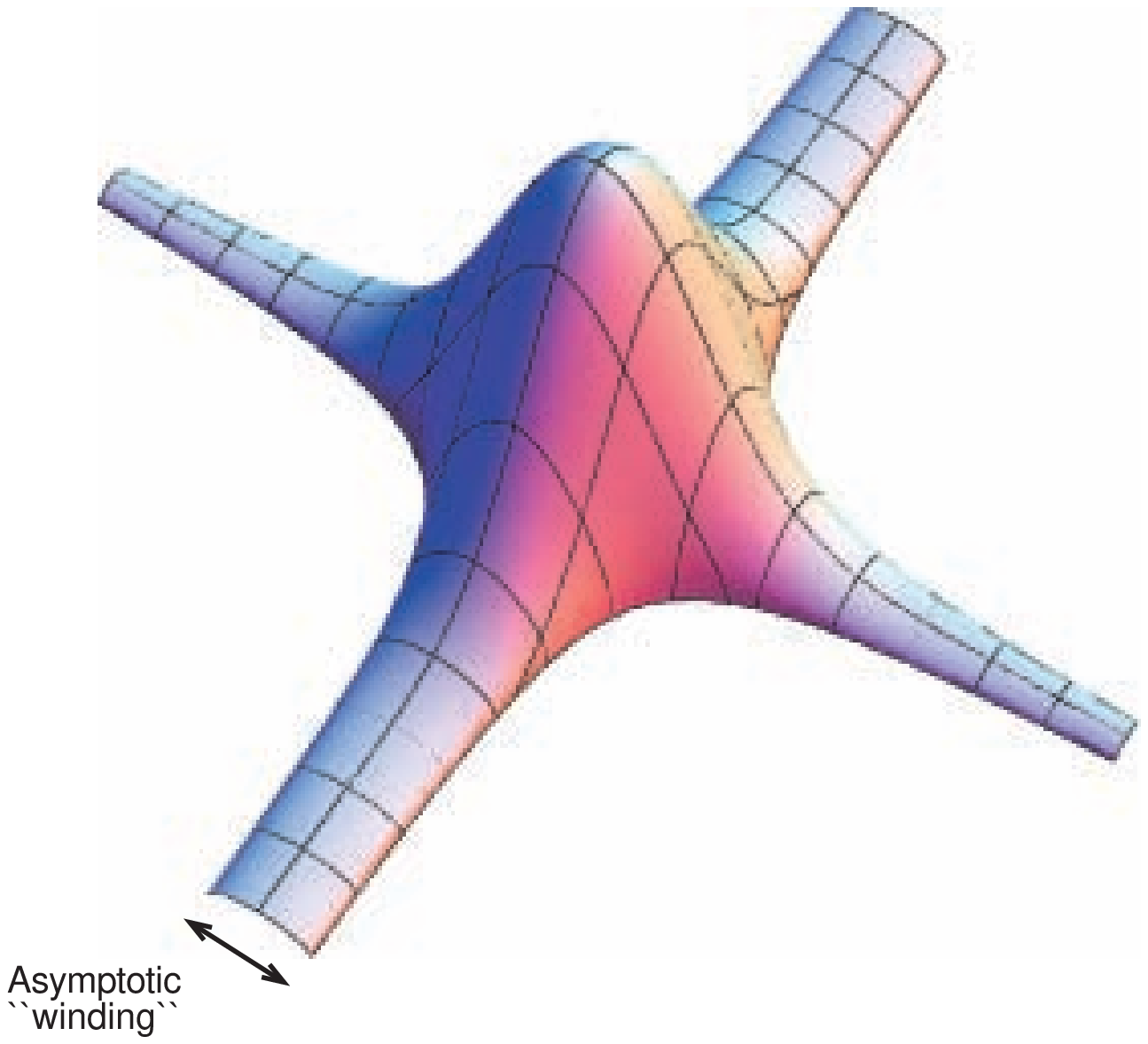}{2.8in}

To see this explicitly, consider first the Gross-Manes flat space
case \refs{\GrossKZ,\GrossGE}. Their solution is
\eqn\grossmanes{\bx=-i\sum_j\bk_jG(z,\sigma_i)~,}
where
\eqn\green{G(z,\sigma)=\log|z-\sigma|^2}
is the Green function in two dimensions and we have parameterized the
disk by the upper half plane $z=\sigma+i\tau$. ``T-dualizing"
\grossmanes, we get
\eqn\Tgrossmanes{\by=-i\sum_j\bk_j\widetilde G(z,\sigma_i)~,}
where
\eqn\dualgreen{\widetilde G(z,\sigma)=\log\({z-\sigma\over \bar
z-\sigma}\)}
is the T-dual Green function. The two Green functions are related by
\eqn\Greenrela{\d_\sigma G=(\d+\bar\d)G=(\d-\bar\d)\widetilde
G=-i\d_\tau\widetilde G~.}
Now, the action evaluated on a solution to the equation of motion is
just a boundary term (when all the vertex operators are inserted on
the boundary):
\eqn\onshell{S=\half\sum_{i\ne
j}\bk_i\cdot\bk_jG(\sigma_i,\sigma_j)~.}
At the extremum, the variation of the action with respect to any of
the open string vertex insertions ($\sigma_i$) must vanish and is given
by
\eqn\actvar{\eqalign{\d_{\sigma_i}S=&\half\bk_i\cdot\sum_j\bk_j\d_{\sigma_i}
G(\sigma_j,\sigma_i)=-{i\over
2}\bk_i\cdot\sum_j\bk_j\d_\tau\widetilde
G(\sigma_j,z)|_{z=\sigma_i}\cr=&\half
\bk_i\cdot\d_\tau\by(z)|_{z=\sigma_i}=0~,}}
where we have used the null relation $\bk_i\cdot\bk_i=0$. Equation
\actvar\ is the Neumann boundary condition on the corresponding
$\by$ segment.\foot{Note that in these singular coordinates, any of
the $\by$ segments is mapped into a single worldsheet point
$\sigma_i$.}

Next, we give a general proof that
extremizing the positions of boundary insertions implies
the Neumann boundary condition. Unlike the previous one, this proof does
not rely on having a flat background, and therefore applies in the
AdS case as well. First note that
\eqn\parti{\d_{\sigma_i}S_{\rm
on-shell}=0\quad\Longleftrightarrow\quad \d_{\sigma_i}e^{-S_{\rm
on-shell}}=0~.}
We are considering only the case where the saddle point
approximation is valid. In such case, to leading approximation and
for any values of the $\sigma$'s, the amplitude is given by the
exponential of the action, evaluated at the saddle point:
\eqn\expct{\<V_1(\bk_1,\sigma_1)V_2(\bk_2,\sigma_2)\dots
V_n(\bk_n,\sigma_n)\> =e^{-S_{\rm on-shell}}~.}
Now, rewrite the left hand side of \expct\ as a path integral with
the action \Buchertwo, which we label as $\widetilde S$. That is,
the path integral, labeled as $\widetilde Z$, is also over the gauge
field $\bA$ and $\by$. At any point in the functional integration, the
fields are integration variables and so their values are some smooth
functions, independent of $\sigma_i$. In that representation (see
\Buchertwo),
\eqn\deriv{\d_{\sigma_i}e^{-\widetilde
S}=i\bk_i\cdot\bA_\sigma(\sigma_i)e^{-\widetilde S}~.}
Plugging this relation into the new path integral (where \Buchertwo\
is the action) and integrating out the gauge field first, we get
\eqn\inthepath{0=\d_{\sigma_i}\log\widetilde
Z=i{\sqrt\lambda\over\widetilde
Z}\,\<\<\bk_i\cdot\bA_\sigma(\sigma_i)\>\>=
-{z_{IR}^2\over\sqrt\lambda}\,
\bk_i\cdot\d_\tau\by(\sigma_i)\quad\Longrightarrow\quad
\bk_i\cdot\d_\tau\by(\sigma_i)=0~.}
The double angle braces $ \vev{\vev{\dots}}$ refer to averages over
$\bf A$ but not $\bf y$. If, on the other hand we integrate out
$\by$ first, then we get
\eqn\inthepatht{0=\d_{\sigma_i}\log\widetilde
Z=i{\sqrt\lambda\over\widetilde
Z}\,\<\<\bk_i\cdot\bA_\sigma(\sigma_i)\>\>=
i\bk_i\cdot\d_\sigma\bx(\sigma_i)\quad\Longrightarrow\quad
\bk_i\cdot\d_\sigma\bx(\sigma_i)=0~,}
which means that the asymptotic open string state created by $V_i$
has zero ``winding". It is important to note that
\inthepath\ and \inthepatht\ are evaluated on the saddle point of
the original path integral
\eqn\orig{\widetilde Z=\<V_1(\bk_1,\sigma_1)V_2(\bk_2,\sigma_2)\dots
V_n(\bk_n,\sigma_n)\>}
and not of the path integral
\eqn\wrongp{\<\<\bk_i\cdot\bA_\sigma(\sigma_i)\>\>=
\<V_1(\bk_1,\sigma_1)V_2(\bk_2,\sigma_2)\dots\d
V(\bk_i,\sigma_i)\dots V_n(\bk_n,\sigma_n)\>~,}
for which the saddle point in question does not contribute.
The condition for extremizing over the positions
of the vertex insertions is the statement that the
descendant field $ \bk \cdot \del \bx e^{i \bk \cdot \bx} $
decouples.

The width of the saddle point is given by
\eqn\saddlet{\eqalign{-\d_{\sigma_j}\d_{\sigma_i}S_{\rm
on-shell}=&i{\sqrt\lambda\over\widetilde
Z}\,\<\<\bk_i\cdot\d_{\sigma_j}\bA_\sigma(\sigma_i)\>\>\cr
=&i\delta_{ji}{\sqrt\lambda\over\widetilde
Z}\,\<\<\bk_i\cdot\d_\sigma\bA_\sigma(\sigma_i)\>\>
=-\delta_{ji}{z_{IR}^2\over\sqrt\lambda}\,
\bk_i\cdot\d_\sigma\d_\tau\by(\sigma_i)~.}}
The leading (divergent) term in $\d_\sigma\d_\tau\by(\sigma_i)$ is
proportional to $\bk_i$ and therefore does not contribute to \saddlet.
Next, we note that the problem has a (dual conformal) symmetry under
which $\by$ and $\bk$ have charge $+1$, whereas $z$ and $z_{IR}$ have
charge $-1$ (so \saddlet\ is invariant). Therefore
\eqn\charges{\d_\sigma\d_\tau\by(\sigma_i)=\sum_l\bk_l\,\wp_{li}
\( { z_{IR}^2\bk_m\cdot\bk_n \over \lambda }\)~,}
where $\wp_{li}$
is some (unknown) function that depends on the cross-ratio of vertex
operator positions.
The inverse width of the saddle is
\eqn\inversewidth{
S'' \sim  { z_{IR}^2\bk_m\cdot\bk_n \over \sqrt \lambda }~
\wp_{ij}\( { z_{IR}^2\bk_m\cdot\bk_n \over \lambda }\).
}
In order to evaluate $\wp$, one needs
some approximate solution near the polygon boundaries.\foot{For the
four gluon amplitude, one may evaluate $\wp$ by studying more
carefully the explicit solutions in \refs{\AldayHR,\MironovQQ}.}
We expect \inversewidth\ to be large in the
regime of interest $\lambda \to \infty, z_{IR} \to \infty$, with
\eqn\limits{1 \ll \lambda^{1/2}  \ll z_{IR}^2 \bk_i\cdot\bk_j \ll
e^{ 2\sqrt \lambda}~~~~~;}
the first limit is for the semiclassical approximation to the $\bx,z$ path integral, the last is to
avoid the large dilaton.  Finally, the indicated behavior of $z_{IR}$ is
the only reasonable regime to study
$ \lambda^{-1}$ corrections or to compare with weak-coupling results.

\subsec{Extremum vs. minimum}

Note that although both the Gross-Manes \GrossGE\ and the
Alday-Maldacena (AM) \AldayHR\ solutions are extrema of the
worldsheet area in the T-dual coordinates, they are not necessarily
{\it minimal} area solutions (not even locally). For the Gross-Manes
four-open-strings amplitude, this can be seen for example by noting
that the second derivative of the on-shell action with respect to
the cross ratio of the vertex insertion point can have different
signs in different physical processes. In the AdS case, this can be
seen by starting with the AM solution for four gluons and rescaling
the radial coordinate ($r$ in \dualPoin), such that the new
embedding extends more deeply into the AdS. The parts of the string
worldsheet near the boundary now become more null and the part where
it closes is pushed farther into the bulk; therefore the area
decreases. This situation is familiar when one uses steepest descent
to approximate a one-dimensional integral in the complex plane. In
that case, there is always one direction along which the saddle
point is a minimum and one along which it is a maximum.

\subsec{Color ordering}

For any given set of $n>3$ gluon external momenta there can be
${(n-1)!\over 2}$ different orderings that are not related by cyclic
permutation or reflection. In AdS, any of these orderings correspond
to a different null polygon boundary condition and therefore to a
different extremal area problem. In addition, the BDS ansatz applies
for any of these orderings.\foot{In AdS, reversing the ordering
of the null segments is obtained by reflecting all the transverse
directions, which is a symmetry. In the field theory, it is the
result of charge conjugation (which may change the overall sign).}
These facts suggest that we should expect to find corresponding
${(n-1)!\over 2}$ different extremal surfaces. We believe that this
is indeed the case, but do not have a proof.

On the other hand, in the flat space case \GrossGE, one can do the
same T-duality transformation and obtain a different extremal area
problem for any ordering. However, in that case, by solving equation
\actvar\ for the four point amplitude one finds that there is an
extremum only for spacelike $s$ and $t$ channel momentum transfer
(only for the $in\to out\to in\to out$ momentum ordering) which, up
to cyclic ordering and reflection, counts as one color ordering.
That is also the only 4-gluon ordering for which we have a known AdS
solution \AldayHR. From the worldsheet point of view, the Chan-Paton
factors generically restrict to a specific vertex operator
ordering, which otherwise would all be part of the same moduli
space. Therefore, one then would be tempted to infer that also in
AdS that should be the case. That is, that for any set of external
momenta, up to cyclic permutations and reflection of the ordering,
there is a unique order for which the string theory path integral in
AdS has an extremum.

As stated above, we believe that the last statement of the previous
paragraph is false, and in this manner, AdS is different from flat
space. That is, in AdS we expect a solution for any ordering. That
point will be important when, in section 5, we will relate the quark
amplitudes to the pure gluon ones.

\subsec{Form factors}

The elaboration of the T-duality transformation in this section has
a possible generalization to the calculation of 'form factors' for
scattering of gluons off of gauge-invariant operators dual to closed
strings\foot{The possibility of
studying such observables in the strong coupling
description was raised in \juanstrings.}. The closed string with nonzero momentum will create a cut
on the T-dual worldsheet and prevents the boundary from forming a
closed loop. One can envision a useful description along these lines
in terms of the universal cover of the embedded string. In the
special case that the closed-string vertex is on-shell, $0=\bk^2
+m^2=q^2$ (so that it has no momentum in the AdS radial direction),
it is pushed (in the classical solution) to the boundary of the
worldsheet, and can be treated classically as an open-string
insertion which does not participate in the Chan-Paton trace. (This
is also what happens in the flat space Gross-Mende calculation with
open strings and a single closed string insertion.) In such a case
one finds the interesting problem of minimizing the action over the
order in which the closed-string insertion appears.

\newsec{The strong coupling dual of planar $\NN=2$ SYM quark scattering}

We add to the $\NN=4$ theory $N_f$ (in general, massive)
$\NN=2$-preserving fundamental fields. At finite $N_f/N$, these
break the conformal symmetry and the resulting theory needs a UV
completion. In the gravity dual, this is described \KarchSH\ by
adding $N_f$ D7-branes that wrap an $S^3\subset S^5$ in the
$AdS_5\times S^5$ geometry. As a result the dual geometry is no
longer asymptotically AdS. However, as long as $N_f
\ll N_c = N$, we can ignore the back-reaction of the D7-branes and
treat them as probes. This is dual to the statement that the field
theory looks conformal, up to a very small scale. The addition of
D7-branes introduces two new open string sectors: the 3-7 strings
which are dual to fundamental hypermultiplet fields, and the 7-7
strings which are dual to operators of the form $\bar q ... q$. The
mass of the hypermultiplets is related to the extent of the 7-branes
in the radial direction (see for example \KruczenskiBE). In this
paper we will consider the case of massless quarks. Therefore, the
D7-branes ``wrap" the whole $AdS_5$.

It is useful to think of the AM prescription as arising from the
near-horizon limit of a flat-space open-string calculation. By
adding D7-branes to this picture, amplitudes with quarks will arise
from the near-horizon limit of disk amplitudes with vertex operators
for 3-7 strings inserted on the boundary. Asymptotic (massless) 3-7
open strings do not extend into the radial AdS direction (or
in any other direction). Therefore, if we fix their dual field theory momenta
then, just as for gluons, their proper momentum is large and the saddle
point approximation will be well-peaked.

Such amplitudes of massless partons are IR divergent and need
regularization. In \AldayHR, two different regularizations were
used. One regularization that is natural from the geometric AdS
point of view, was to hold the branes associated with the color
indices of the scattered gluons at some fixed radial position
$z_{IR}$. This provides an IR cutoff on the calculation. This cutoff
also regulates the quark amplitudes (as will soon be obvious). A
second regularization used in \AldayHR\ is dimensional
regularization. It is achieved by replacing the D3-branes by
Dp-branes, where $p=3-2\epsilon$. It could be extended to the case at
hand by replacing the D7-branes with D$\tilde p$-branes, where
$\tilde p=6-2\epsilon$; we do not pursue this possibility here.

\subsec{Boundary conditions at the D7 cusp}

To find the saddle point solution, it will again be useful to study
the `T-dual' description. The T-duality is done only along
directions shared by the D3-branes and D7-branes. Each of the vertex
operators has a $e^{i\bk\cdot\bx}$ factor, which causes the T-dual
$\by$ coordinate to jump by $\bk$ at the insertion point. Along the
components of the boundary between the vertex operators, the image
in the target space lies at a fixed value of $\by$: this is T-dual
to the statement that the $\bx$ coordinates of the worldsheet
satisfy Neumann boundary conditions when ending on a D3-brane:
$$ 0= \del_\tau\bx |_{\del \Sigma}~ \propto~ \del_\sigma\by |_{\del \Sigma} .$$
Therefore, the projection of the worldsheet into the Minkowski space
is again a polygon specified by the momenta, with the order
determined by the Chan-Paton ordering.
\fig{The worldsheet for $\bar qggq$ scattering, and its image in the
T-dual $AdS$.  The quark and antiquark vertex operators change
boundary conditions from the 3-branes to the 7-branes and
back.}{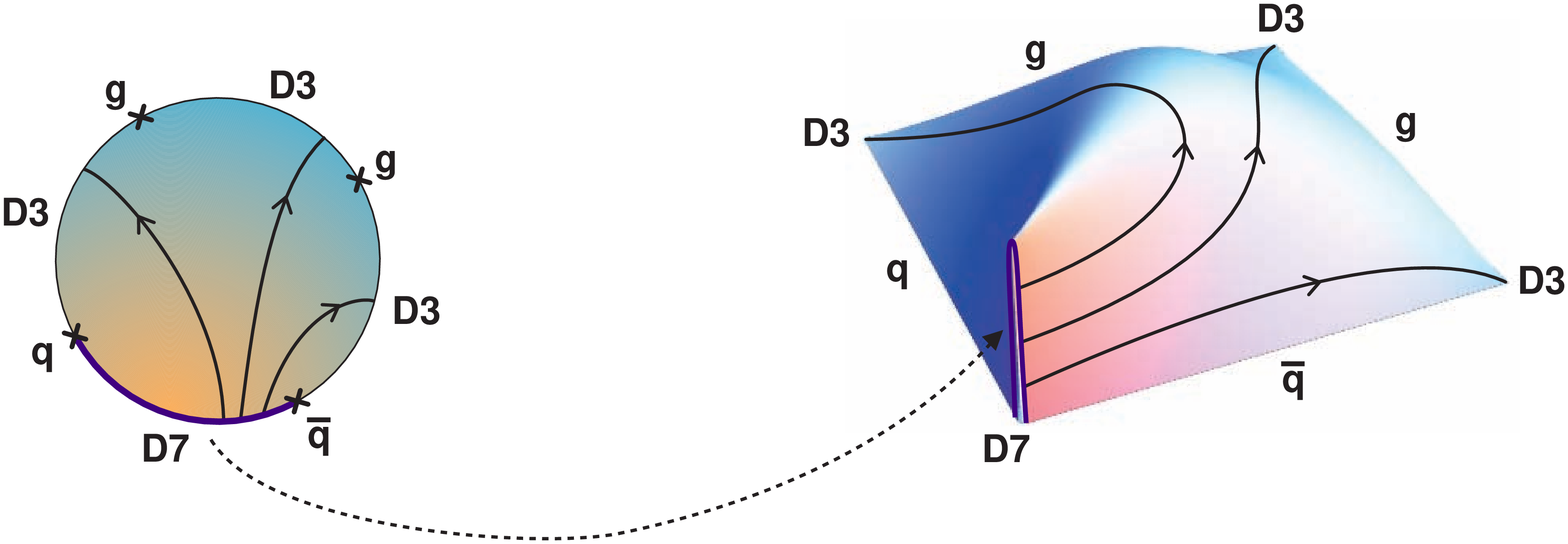}{5in}
%
As in the pure gluon amplitude, the components of the boundary
ending on the D3-branes have Dirichlet boundary conditions in the
radial direction ($r$). When the string ends on the D7-brane,
however, it satisfies Neumann boundary conditions in the radial
direction, and can extend into the bulk. Since this component of the
boundary lies at a fixed value of $y$, such an extension must fold
back on itself (see Fig.\ 3).\foot{The possibility that the
worldsheet could end on a folded string was independently recognized
in \KomargodskiER\ with, however, a contradicting conclusion.}

In summary, the inclusion of the quark-antiquark pair amplitude
introduces a new kind of cusp to the light-like polygon Wilson loop,
above which the worldsheet {\it may} end on a folded string. Next, we will
see that it does.

\newsec{Relating planar gluon scattering to quark scattering at strong coupling}

AdS solutions for quark and gluon-quark amplitudes can be constructed
from special gluon amplitudes. Specifically, consider a worldsheet
ending on a light-like Wilson loop with a {\it self-crossing}, as
follows.

For definiteness, we focus first on the $\bar qggq$ (anti-quark,
gluon, gluon, quark) amplitude. Consider the polygon associated to
6-gluon scattering with color-ordered momenta satisfying $\bp_6 =
\bp_2$, $\bp_5 = \bp_3$. The conservation of momentum can now be
written as
\eqn\crossingcondition{ \half \bp_1 +\bp_2 + \bp_3 + \half \bp_4 =
0~.}
The resulting polygon crosses itself at the midpoints of the lines
associated with $\bp_1$ and $\bp_4$ (see Fig.\ 4). Let $\by=0$ be
the point of crossing.
\fig{The polygon associated to orientifold-symmetric 6-gluon
scattering.}{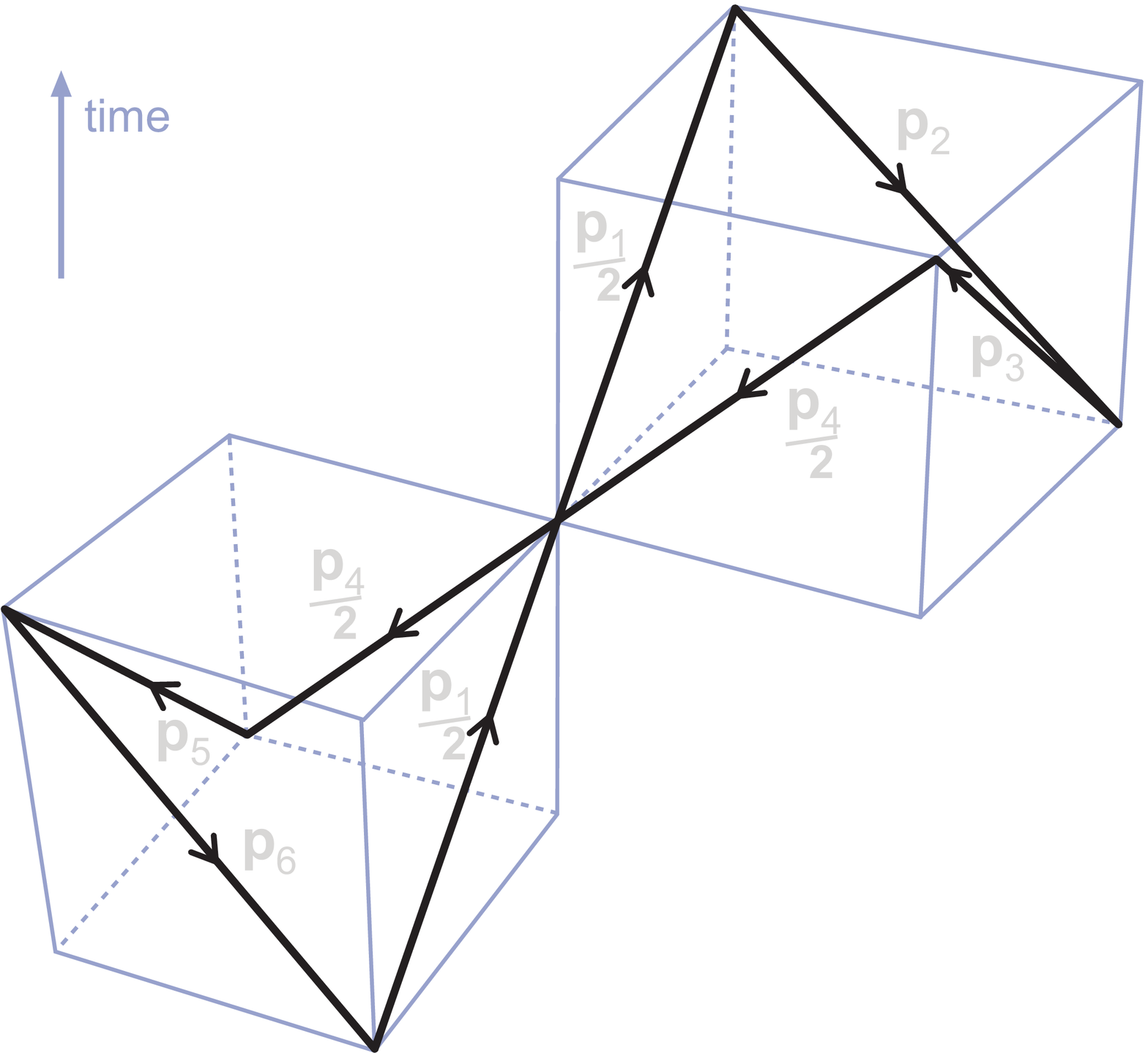}{2.6in}
The polygon is mapped to itself by a symmetry which reflects through
the crossing point, while simultaneously reversing the orientation
of the gluon lines. Assuming there is a unique worldsheet which
extremizes the area with these boundary conditions (see section
3.4), it too is mapped to itself by $R\Omega$, where
\eqn\reflection{R:\qquad (\by,r)\to(-\by,r)~,}
extends the action on the momenta to the whole T-dual AdS space, and
\eqn\orientifoldcondition{ \Omega:\qquad(\sigma,\tau) ~\to~
(-\sigma, \tau)}
is the worldsheet parity; we have parameterized the disk by the
upper half plane coordinates $(\sigma,\tau\ge 0)$.\foot{Note that
$R\Omega$ acts in the same way on the $(\bx,z)$-space \Poincare.}

The six-gluon worldsheet is not yet known. Nevertheless, we can
infer the following from the $R\Omega$ symmetry. Consider the two
boundary points $\sigma_1=0$ and $\sigma_4=\infty$ where the
corresponding vertex operators are inserted. There is a curve on the
worldsheet $\gamma=(0,\tau)$ which connects them (see Fig.\ 5), and which is made
of points that are invariant under $\Omega$. If we assume that the
radial direction varies smoothly as we cross $\gamma$, then it
follows from the $R\Omega$ symmetry that it satisfies Neumann
boundary conditions there.\foot{Note that from the worldsheet point
of view, {\it at} the boundary, nether $r$ or $z=1/r$ are good
coordinates and one had better use the coordinate $\phi=\log(z)$.}
\fig{Instructions for cutting.}{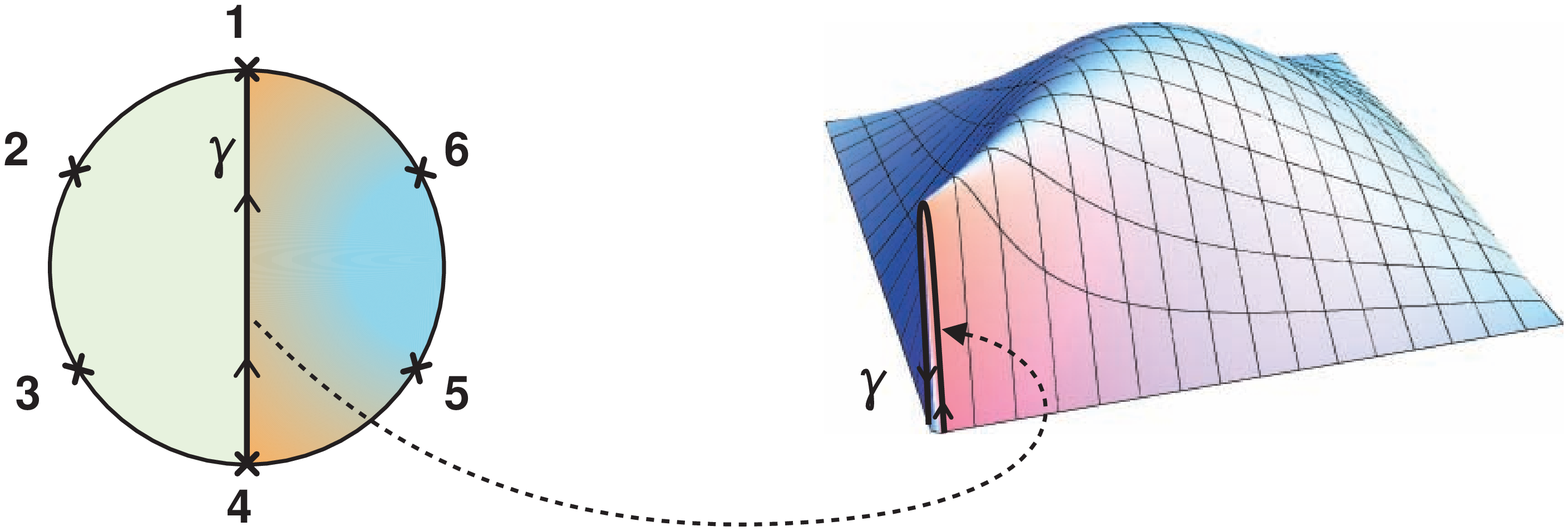}{5in}

To relate this picture to the quark amplitude, it is crucial that
the string does not ``open up" above the crossing point. That is,
we require
that the image of $\gamma$ is a folded string at $\by=0$. To see
that this is indeed the case, note that modding out by $R\Omega$ is
equivalent to placing an orientifold line that extends into the
radial direction above that point ($\by=0$). In that picture, our
open string worldsheet is one-half of a closed string worldsheet with crosscap.

So far, we have argued that the extent of the folded string
along the ``orientifold line", $l$, can vary from $0$ to $\infty$ in the
configuration space. The following argument is evidence that $l>0$.
One half of the divergent piece of the self-crossing 6-gluon
amplitude is smaller then the divergent piece in the 4-gluon
amplitude. If the solution did not extend into the radial direction
above the crossing point ($l$=0), it would satisfy the AM boundary
conditions as well. If such a solution existed and, for the 4-gluon
amplitude, were different from the AM solution, AM should have
summed over it as well, and it would dominate over the contribution
of their solution! But the contribution of their solution, by
itself, agrees with the field theory prediction. Note that the AM
four-gluon solution (explicitly given in \AldayHR) does not satisfy
Neumann boundary conditions in the radial direction at the D3
boundaries (D3-cusps).

If we cut the worldsheet along $\gamma$, each identical half is the
desired worldsheet describing the $\bar qggq$ scattering with quark
momenta $\bk_1 = \half \bp_1$ and $\bk_4 = \half \bp_4$, and gluon
momenta $\bp_2, \bp_3$. The configuration space for the $\bar qggq$
worldsheet (made of all embeddings invariant under $R\Omega$) is a
subspace of the one for the six-gluon worldsheet with crossings.
Therefore, an extremum of the worldsheet path integral with 6-gluon
boundary conditions that is $R\Omega$-symmetric, is in particular an
extremum of the worldsheet path integral with the quark-scattering
boundary conditions.\foot{Note that the reverse of this statement is
not true. Here we rely on the assumption that there is a unique
extremal surface in AdS corresponding to the ordering with crossing
(see section 3.4).}

By $R\Omega$-symmetry, the area is half the area of the six-gluon
worldsheet. The value of the $\log$ of the $\bar qggq$ amplitude
($\SS=-\log(\AA)$) in the large-$\lambda$ planar limit, then, is
half the value of the $\log$ of the six-gluon amplitude (see
Appendix 1 for notations):
%
\eqn\mainresult{\eqalign{&-\SS_{qgg\bar q}(k_1, k_2, k_3, k_4) =
-\half \SS_{6g} (2k_1, k_2, k_3, 2 k_4, k_3, k_2)\cr\cr
=&-{f(\lambda)\over 8}\[\ln^2\({\mu^2\over -2s}\)+
\half\ln^2\({\mu^2\over -t}\)-\half \ln^2\({4s\over -t}\)
+\li2\(1+{s\over t}\)- {9\over 2}\zeta_2\]\cr\cr &-{g(\lambda)\over
2}\[\ln\({\mu^2\over -2s}\)+\half\ln\({\mu^2\over -t}\)\]-{1\over
4}h(\lambda)~,}}
where we have used the explicit expression for the finite part of
the 6-gluon amplitude given in Appendix A and $s=-(\bk_1+\bk_2)^2$,
$t=-(\bk_2+\bk_3)^2$.\foot{We remind the reader that in this paper
we work with $(-+++)$ signature.} To get \mainresult, one has to use the
relation $\tn{n}{i}=\tn{6-n}{3-i}$. It follows from the $R\Omega$
symmetry where $i\to (3-i)$ implement the reflection and $n\to
(6-n)$ implement the orientation change.\foot{Note that always
$\tn{n}{i}=\tn{6-n}{n+i}$. Combining this relation with our symmetry
gives $\tn{n}{i}=\tn{n}{3-n-i}$.}
\fig{The 6-gluon worldsheet embedding passes through itself above
the crossing. In doing so, it reverses orientation (indicated by
color). Near the crossing, the embedding (in $\IR^3$) is called a
`Whitney umbrella'.}{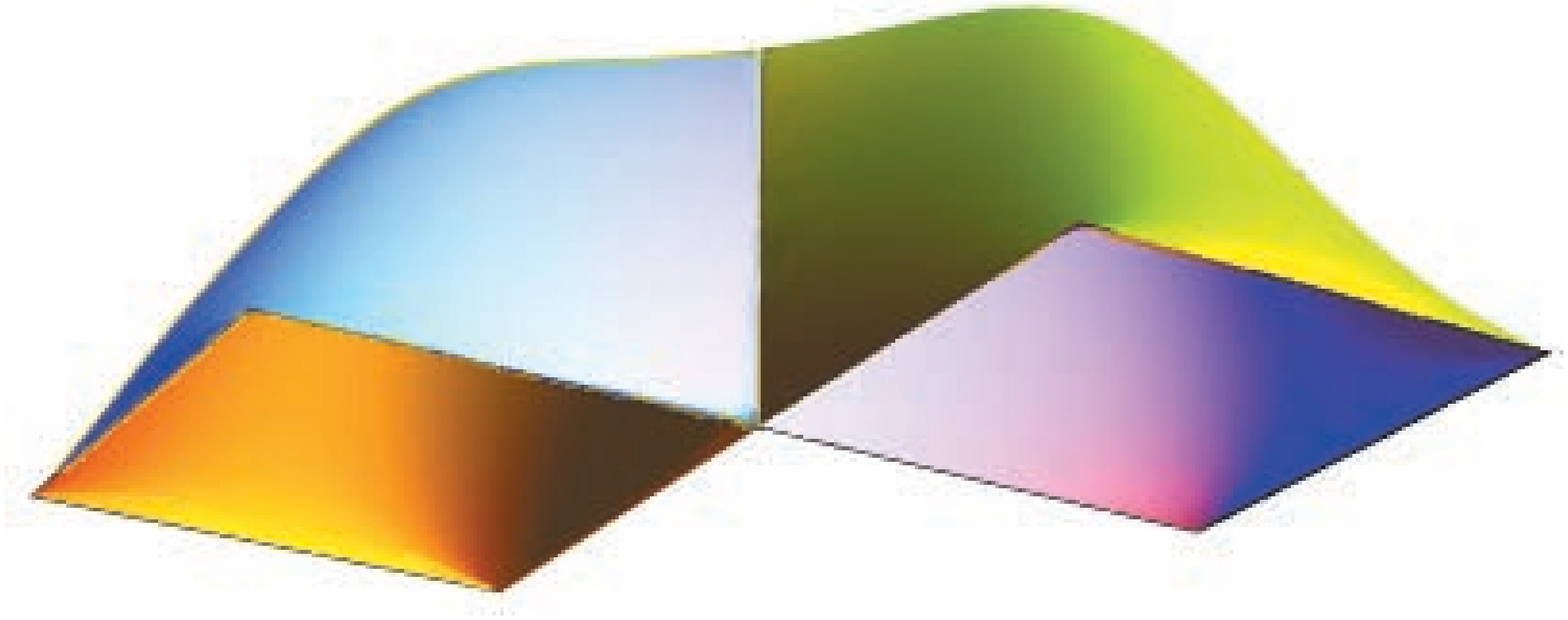}{3.5in}

The image of $\gamma$ in the target space is the folded string. We
expect the maximum value of $r$ along $\gamma$ to be of order
$\sqrt{s_{q,\bar q}}$, where $s_{q,\bar q}=-(\bk_q+\bk_{\bar q})^2$
is the Mandelstam variable associated with the adjacent quark and
anti-quark ($s_{14}$ in the example above). The intuition is that it
is advantageous in extremizing the area for the interior of the
worldsheet to extend into the bulk of AdS, and this suspends the tip
of the fold away from the boundary. Therefore the extension of the
folded string into the radial direction is expected to be of order
$\sqrt{-\(\sum_i\bk_i\)^2}$, where $i$ runs over all momenta in the
loop except the ones of the two adjacent quarks in question (so
$\sum_i\bk_i=-(\bk_q+\bk_{\bar q})$).

The six-gluon worldsheet that we have described has a
singularity.\foot{The Whitney umbrella is a {\it stable} singularity
of a map from $\IR^2 \to \IR^3$; {\it stable} here means stable with
respect to variations preserving the {\it $k$-jet} (for some $k$) of
the singularity. Our solution seems to provide an interesting
connection between jets in physics and jets in mathematics.} One can
worry that such a development calls into question our classical
description.
The singular point is measure zero in the integral that produces the
area; its only effect can come when we consider the variation of the
action under small fluctuations around the extremum. The specific
concern is that there could be a mode localized at the singularity
which is a zeromode of the fluctuation matrix $S''$.\foot{Note that
any non-zero eigenvalue of $S''$ is proportional to $\sqrt\lambda$.}
Since the singularity is stable, any zero mode that is localized at
the singularity can only change its location but cannot smooth it
out. One such mode is the mode that changes the height of the fold
($l$ above). If we focus on the D7-cusp (along the lines of
\KruczenskiFB), we find that $l$ is not determined locally.\foot{We
thank Hong Liu for discussions of this point.} We expect the tension
of the rest of the worldsheet provides a potential for it
(proportional to $\sqrt\lambda$). The other dangerous modes
change the location of the singularity in the four
transverse directions (and are odd under $R\Omega$). Without the
knowledge of the explicit solution we cannot determine their fate.
Here we assume such zeromodes are absent.


Note that it is crucial for our picture that the crossing is along
null lines. In \DrukkerZQ, it is shown via AdS/CFT that the
expectation value of an Euclidean Wilson loop with a self-crossing
is equal to the sum of the expectation values of the two sub-loops.
This result follows from the infinite area cost for the Euclidian
string to extend near the boundary. Therefore, cutting the Euclidian
solution at some $r=\epsilon$ away from the boundary divides it into
two components (see figure 7 of \DrukkerZQ), each of which is the
extremal worldsheet for one of the sub-loops.

In Minkowski space however, it does not cost area for the worldsheet
to extend along a null surface - even if it is near the AdS boundary.
Therefore, it needn't be true that the worldsheet pinches to the
boundary at the crossing. Indeed, we have given evidence above that
in our case of interest it does not. (In the next section we will
further verify
this in the dual field theory.)  In such a case, slicing the
worldsheet at some small distance from the boundary does {\it not}
separate it into two disjoint Wilson loops (see Figs.\ 3 and 5).

As for the AM case, here, in the T-dual picture, we can ignore the
dilaton unless the worldsheet extends into the region where it blows
up. The dilaton admits its maximal value at the cutoff
($r_{UV}=\sqrt\lambda/z_{IR}$) and therefore, the presence of the D7
cusp does not invalidate that approximation.

\subsec{Generalizations}

By considering other symmetric gluon scattering amplitudes,
we can construct the generic amplitude with quarks.

The next simplest case is two quarks and any number of gluons.
Specifically, consider a scattering amplitude with two quarks with
momenta $\bk_1$, $\bk_n$, and $n-2$ gluons with momenta
$\bk_2,~\dots~\bk_{n-1}$. We will study an auxiliary $(2n-2)$-gluon
scattering process with a $\IZ_2$ symmetry $R\Omega$. Label the gluon
momenta
$$\bp_1 = 2\bk_1~,\qquad\bp_{n} = 2\bk_n~,\qquad\bp_{n+l}
= \bp_{n-l} = \bk_{n-l}~,\quad l= 1,~\dots~n-2~.$$
Momentum conservation of the $\bar q g^{n-2}q$ process
($\sum_{i=1}^n \bk_i = 0$) means that the gluon momenta satisfy the
condition
\eqn\ncrossingcondition{ \half \bp_1 + \bp_2 +~ \dots~ +\bp_{n-1} +
\half \bp_n = 0,
}
which says that the $\bp_1$ line and the $\bp_n$ line crosses at
their midpoints. The resulting polygon is $R\Omega$-symmetric and
therefore, we expect the extremal worldsheet ending on it to be
$R\Omega$-symmetric as well. The area for the $(2n-2)$-gluon
amplitude is twice the amplitude for $\bar q g^{n-2}q$:
\eqn\mainresultnified{ \SS_{\bar qg^{n-2}q}(\bk_1, \bk_2,\dots,
\bk_{n-1}, \bk_n) = \half \SS_{g^{2n-2}} (2\bk_1, \bk_2,\dots,
\bk_{n-1}, 2\bk_{n}, \bk_{n-1}, \bk_{n-2},\dots, \bk_2)~.}

To construct an amplitude with more than one pair of quarks, proceed
as follows. Draw the closed polygon associated with the
color-ordered momenta of the quarks and gluons in question.
The quarks must come in quark anti-quark adjacent
pairs.\foot{The possible exception to this is the insertion of 7-7
strings between the quarks.
According the the gauge/string dictionary,
such states are dual to mesons.  The corresponding
amplitudes then describe `form factors' for
the scattering of quarks and gluons off of these mesons.
We leave their study for future work.
}
The
quark lines in adjacent pairs meet at a D7-cusp like the one
described in detail above. At each of these D7-cusps, append the
image of the polygon under the $\IZ_2$ symmetry which acts locally
as described above. As
soon as there is more than one pair of quarks, there will be
infinitely many images (just as when mirrors face each other). We
can regulate this by restricting the images to some finite volume
$V$, and ending the graph in some way which breaks the symmetry but closes the
color trace. Because this is a polygon in flat space, at large
volume $V$, the contributions from the parts of the polygon near the
boundary are negligible.

\fig{Prescription for the $q\bar q q\bar q$
amplitude.}{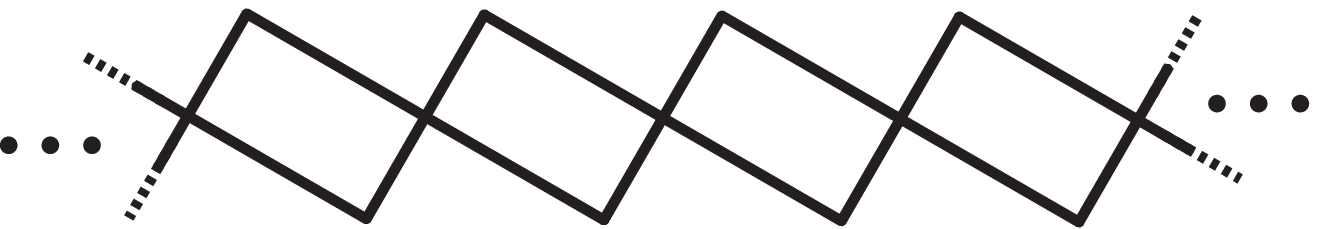}{3.5in}
For example, we can construct the amplitude for $ q \bar q \to q
\bar q$ by considering the process with infinitely many gluons
indicated in Fig.\ 7.
This leads to the relation
\eqn\mainresultnifiedmore{\eqalign{M_{\bar qq\bar qq}(\bk_1, \bk_2,
\bk_3, &\bk_4)\cr =\lim_{\ell \to \infty} {1\over \ell}  M_{g^{2
\ell}} (\bk_1, &\underbrace{2\bk_2, 2\bk_1, 2\bk_2,~\dots~,2\bk_1},
\bk_2, \bk_3, \underbrace{2\bk_4, 2\bk_3, 2\bk_4,~\dots~,2\bk_3},
\bk_4)~.  \cr &~~~~~~~~~~\ell-2~{\rm
times}~~~~~~~~~~~~~~~~~~~~~~~~~\ell-2~{\rm times}~~~~~~~~ }}

\newsec{No Sudakov form factors from D7 cusps}

In the previous section we saw that the extremal area surfaces
corresponding to quark scattering can be obtained from gluon ones
with self-crossing. In this section we study the implications of
this statement in field theory. We will first show that in one-loop
gluon amplitudes, no new divergence arises from such a crossing.
Given the structure of the BDS ansatz, this implies that no new
divergence arises to all orders. Next, we give a pure field theory
derivation of this prediction, by showing that the Sudakov form
factors corresponding to D7 cusps are down by $1/N$ and are
therefore absent in the strict large $N$ limit. This supports
the claim of \S5 that
on top of the crossing point,
there is a folded string which extends in the radial direction, as
drawn in Fig.\ 3.

\subsec{One-loop gluon amplitude with crossing}

As far as the one loop gluon integral is concerned, no new
divergence arises when the loop made of the successive {\it null}
external momenta develops a self-crossing. This can be shown by
directly studying the $\NN=4$ six-gluon one-loop amplitude
 (which reduces to a scalar box
integral, see \eg\ \DrummondAU\ and references therein). An intuitive
way to extract the contribution to the one-loop integral from the
crossing is to study its ``double loop" representation \dloop\
\refs{\DrummondAU, \BrandhuberYX}
reviewed in \S2. In that representation, divergences arise when the
two integrals run over adjacent legs of the polygon. As a
preparation, consider first the contribution to \dloop\ from the
region of integration where both $\by$ and $\by'$ are near a cusp
made of two successive momenta $\bk_1$ and $\bk_2$. We parameterize
$\by$ and $\by'$ as
\eqn\yparm{\by=-\tau\bk_1~,\qquad\by'=\tau'\bk_2~,}
where $\tau$ and $\tau'$ run from 0 to 1. The part of the integral
\dloop\ where $\by$ runs on the $\bk_1$ segment and $\by'$ runs over
the $\bk_2$ segment is:
\eqn\crossing{\eqalign{\mu^{2\epsilon}\int_{-\bk_1}^0\int_0^{\bk_2}
\frac{d\by\cdot d\by'} {[-(\by-\by')^2]^{1+\epsilon}}=&\half
\({\mu^2\over-2\bk_1\cdot\bk_2}\)^\epsilon\int_1^0{d\tau\over\tau^{1+\epsilon}}
\int_0^1{d\tau'\over\tau'^{1+\epsilon}}\cr\cr =&-\half{1\over
\epsilon^2}\({\mu^2\over-s_{1,2}}\)^\epsilon~,}}
where it was crucial that $\epsilon<0$ for \crossing\ to converge.

Now consider a loop $\Pi_\infty$ made of successive {\it null} segments that has
a single crossing. It can be thought of as two loops $\Pi_{1,2}$
touching at a cusp point with completing angles and opposite
orientations, $\Pi_\infty=\Pi_1\cup\Pi_2$. The double loop integral
\dloop\ now takes the form
\eqn\crossl{\eqalign{M^{(n)}_1=&
\half\mu^{2\epsilon}\ooint_{\Pi_1\cup\Pi_2}\ooint_{\Pi_1\cup\Pi_2}
\frac{d\by\cdot d\by'} {[-(\by-\by')^2]^{1+\epsilon}}\cr\cr=&
\half\mu^{2\epsilon}\(\uoint_{\Pi_1}\uoint_{\Pi_1}+
\doint_{\Pi_2}\doint_{\Pi_2}+2\uoint_{\Pi_1}\doint_{\Pi_2}\)
\frac{d\by\cdot d\by'} {[-(\by-\by')^2]^{1+\epsilon}}~.}}
The first two double loop integrals in the last line of \crossl\ are
just \dloop\ evaluated on the loops $\Pi_1$ and $\Pi_2$. If that
were the whole answer one would conclude that the intersection gave
double the divergent contribution \crossing\ (which after expanding
in powers of $\epsilon$ are the one-loop contribution to the Sudakov
form factors). However, the double cross-loop integrals in the last
line of \crossl\ exactly cancel these divergences. To see this
cancellation, we simply include the cross-terms in \crossl\ by
allowing
$\tau$ and $\tau'$ in \crossing\ to proceed beyond the zero:
\eqn\tcross{\int_1^0{d\tau\over\tau^{1+\epsilon}}
\int_0^1{d\tau'\over\tau'^{1+\epsilon}}
\quad\to\quad\int_1^{-1}{d\tau\over\tau^{1+\epsilon}}
\int_{-1}^1{d\tau'\over\tau'^{1+\epsilon}}=
\[{1-(-1)^{-\epsilon}\over\epsilon}\]^2}
%
which is finite as $\epsilon\to 0$. In some abused language, it is
the presence of these cross terms between the two loops that causes the worldsheet
to extend into the radial direction above the crossing point.

\subsec{Sudakov and the $1/N$ expansion}

A colored particle participating in a scattering process is very
unlikely not to emit soft gluons. In the perturbative calculation of
exclusive amplitudes, this statement manifests itself in the form of
an IR divergent exponential suppression factor. In
physical quantities,
these divergences are replaced by and
determine important dependence on the definition of the observable
(detector sensitivity, cone angles) \StermanWJ. These divergences can be
resummed into a `Sudakov factor' of the form
\refs{\sudakovreferences,\Korchsudakov,\sudakovreview,\MagneaZB,\catani,\bds}
(and see \AldayMF\ for a recent discussion in this context),
\eqn\resummed{\AA\sim
e^{-h(\lambda)\log^2(\mu_{IR})-h'(\lambda)\log(\mu_{IR})}~,}
where $h, h'$ are some functions of the coupling. The form of the
soft gluon exchange diagrams that contribute to the Sudakov form
factor is shown in Fig.\ 8.
\fig{An example of the soft gluon exchange diagrams that resum into
the Sudakov form factor. The curled lines are gluons, where the
external line from which the soft gluons are emitted represent any
colored particles.}{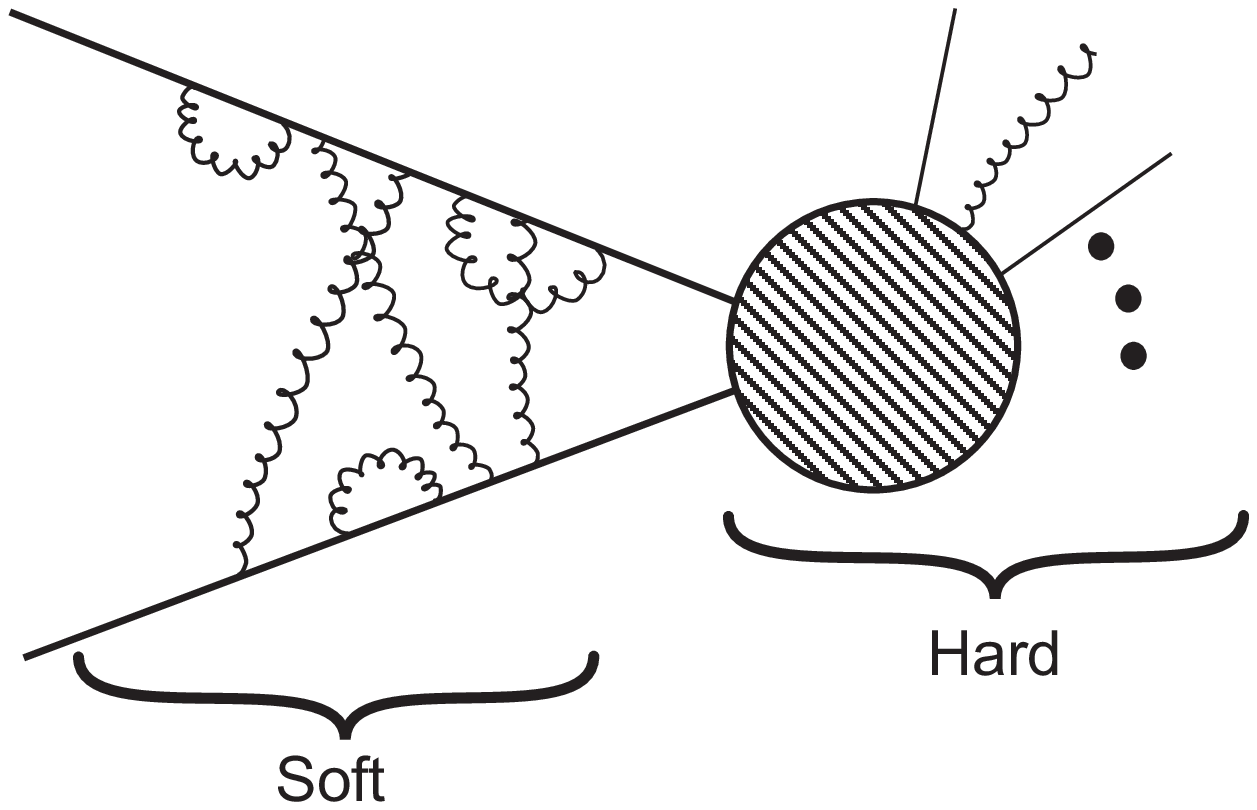}{3in}

For massless external particles,
the Sudakov form factor looks like
\eqn\masslesssudoku{
e^{-{f(\lambda)\over 4}\log^2(\mu)} }
as a function of the IR cutoff
$\mu$
\refs{\sudakovreferences,\Korchsudakov,\MagneaZB,\catani,\AldayMF}
(for a review see \sudakovreview), where $f(\lambda)$ is the cusp
anomalous dimension. This factor appears in front of the {\it whole}
amplitude, it gives the leading IR behavior when we consider the
exclusive scattering of colored particles, and it is an important
ingredient in the computation of amplitudes \CollinsGX.

In a planar diagram contributing to gluon scattering, the left index
line from each gluon ends at the right index line of the next. For
each consecutive pair we get a factor of the form \resummed, with
the function $h$ given by $f/4$. An example of such a planar soft
gluon exchange diagram is drawn in Fig.\ 9 in the double line
notation.
\fig{An example of a planar soft 4-gluon exchange diagram that
contribute to the corresponding Sudakov form
factor.}{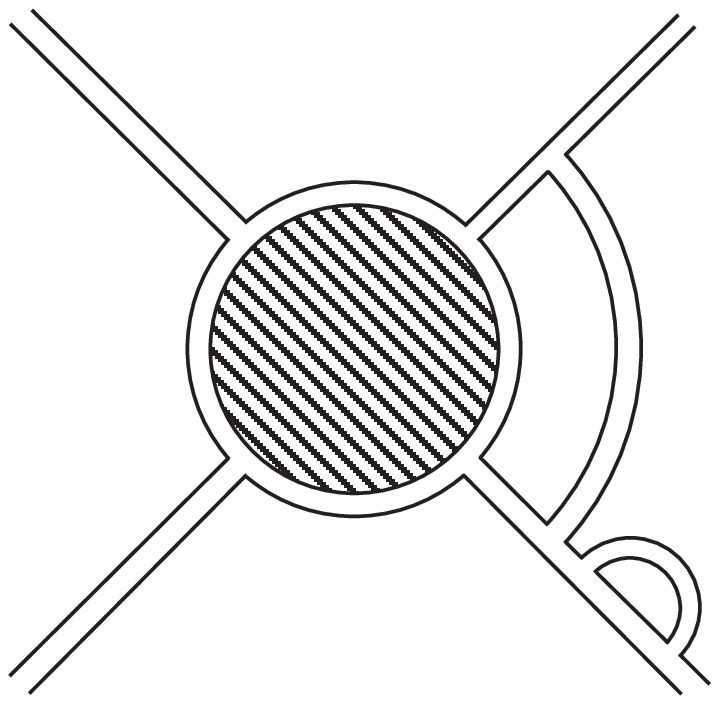}{1.6in}

Imagine now that we replace two successive gluons by quark
anti-quark pair. As can be seen in Fig.\ 10, a gluon exchange
between the quark and anti-quark results in a hole in the worldsheet
and is therefore down by $1/N$.
\fig{An example of a quark anti-quark to 2-gluon diagram where the
quark and anti-quark exchange soft gluon. It has an annulus topology
and is therefore down by $1/N$ with respect to the same diagram
without the soft gluon exchange.}{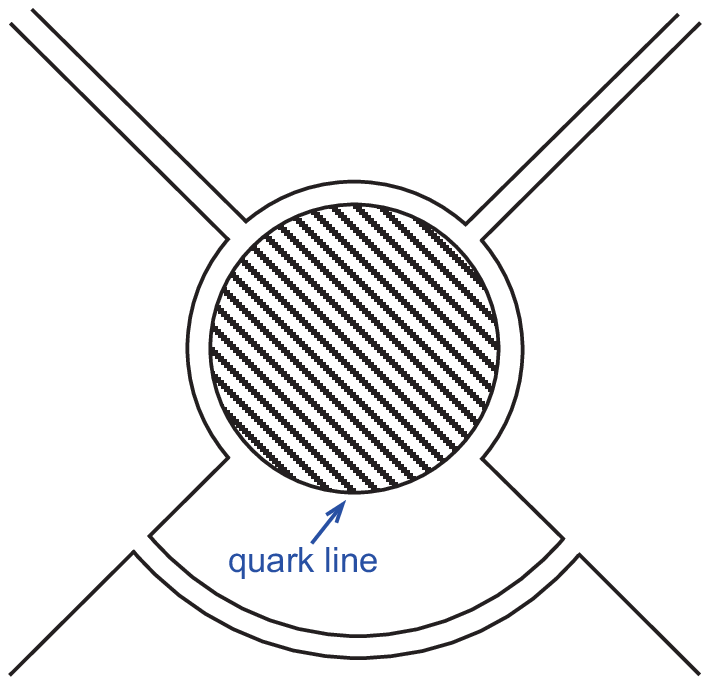}{1.6in}
This does not mean that there is no Sudakov form factor from soft
gluon exchange between the quark and the anti-quark. All it means is
that these diagrams combine with the non-planar diagrams so that the
coefficient in front of the $\log^2(\mu)$ in the Sudakov form factor
scales as $1/N$ and disappears if we take the large $N$ limit while
holding the IR cutoff ($\mu$) fixed. Note that any external quark
does lead to a Sudakov form factor that is not down by $1/N$. It
results from exchanging soft gluons with the next hard quark or
gluon {\it along the same color line}, as in Fig.\ 11. This
divergence, however, corresponds to a D3-cusp in the dual string
picture.
\fig{An example of a planar quark anti-quark to quark anti-quark
diagram where successive quark anti-quark exchange a soft gluon. It
contribute to the corresponding Sudakov form factor and is not down
by $1/N$.}{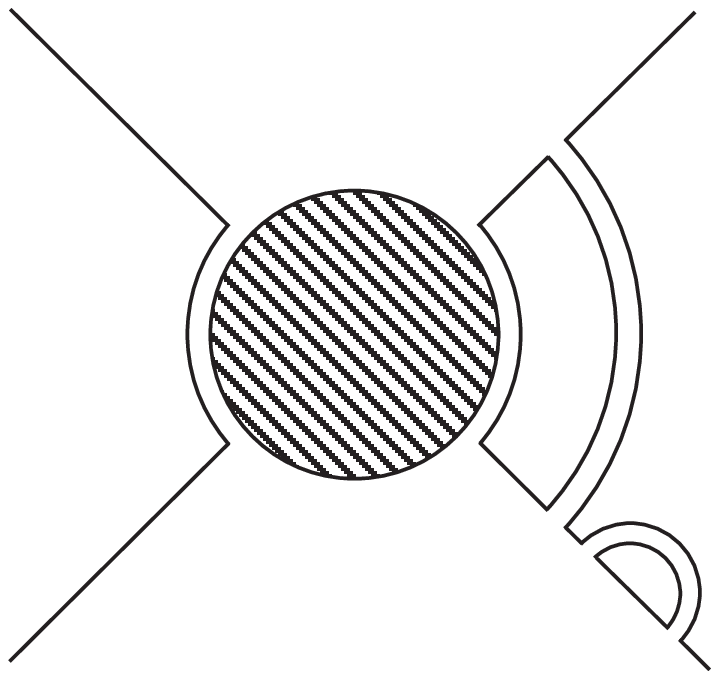}{1.6in}

A limit of the AM construction is the statement that the Sudakov
double logs can be computed by replacing the hard partons by Wilson
loops \refs{\Korchsudakov, \sceft}. The factor in front of the
double logarithm ($f(\lambda)$, the cusp anomalous dimension) for a
Wilson line in the fundamental representation is half the one for a
Wilson line in the adjoint representation. That factor of half is
exactly the manifestation of what we have just seen. As was recently
explained in \AldayMF, the coefficient $f(\lambda)$ in the Sudakov
form factor is the tension of the flux tube that is stretched from
the colored particle. The fact that for a Wilson line in the adjoint
representation this factor is twice the one for a Wilson line in the
fundamental representation can be interpreted as saying that there
are two flux tubes stretching out of a particle in the adjoint
representation, one to the `right' and one to the `left' with
respect to the planar structure of the amplitude (see Fig.\ 12). In
the Wilson line picture, they are folded on top of each other and
are equivalent to a single flux tube with doubled tension.
\fig{An example of planar diagram that contributes to the flux tube
extended from Wilson line with cusp. {\bf a.} Wilson loop in the
adjoint has two flux tubes. {\bf b.} Wilson loop in the fundamental
has one flux tube.}{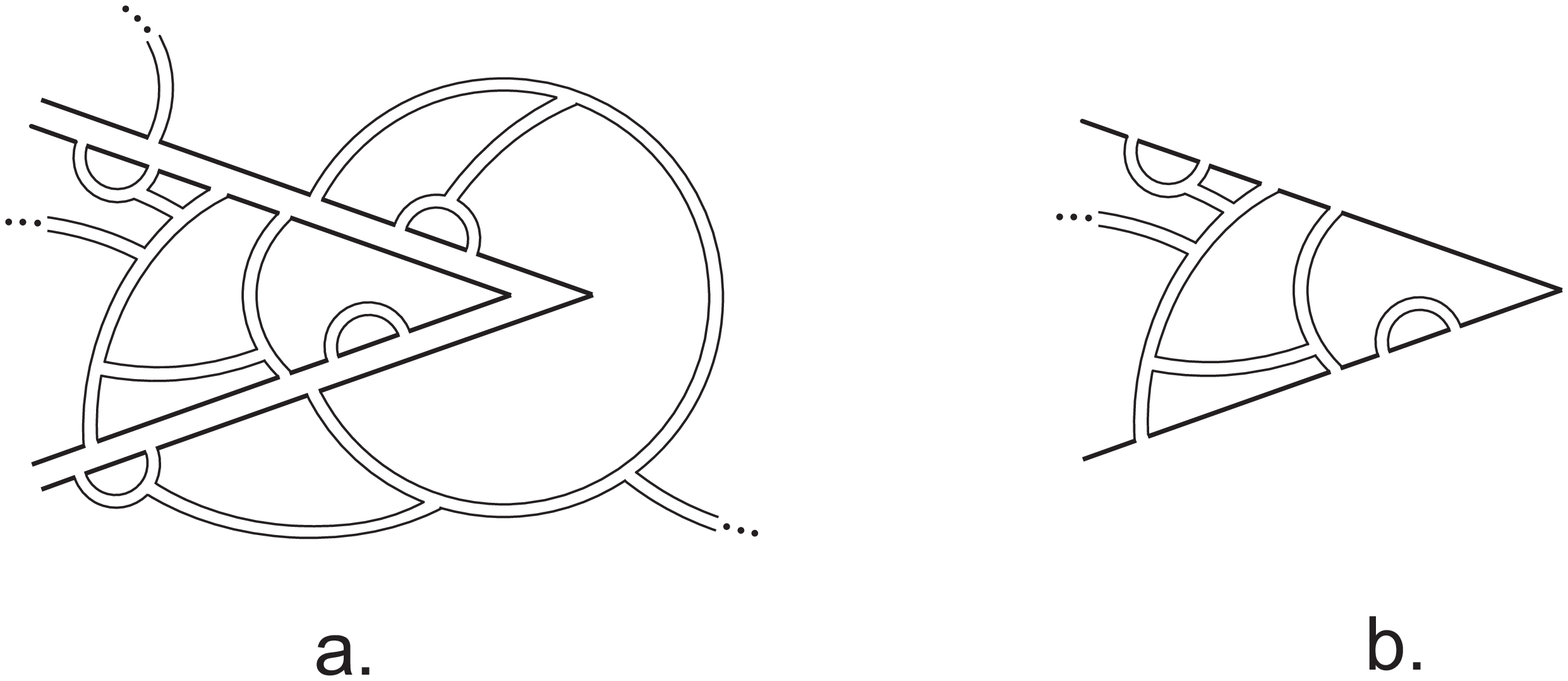}{4in}

\newsec{Comparison with perturbative results}

\subsec{Non-BDS for quark amplitudes}

As was checked up to 4-loops (see \BernCT\ for 5-loops ansatz) and
conjectured by BDS \bds, the planar MHV one-loop gluon amplitude in
$N=4$ SYM exponentiates (see section 2). One might naively expect
that a similar exponentiation holds for planar HMV one-loop massless
quark amplitudes in $\NN=2$ SYM (in the probe approximation). Such a
relation would make our conjecture easy to check. However, this
cannot be the case. In the pure gluon amplitude, one would
picturesquely expect that when expanding the exponent of the
one-loop amplitude, many such one-loop amplitudes will ``tile''
together to form the string worldsheet. In the t' Hooft large $N$
expansion of the quark-gluon amplitude, the quark propagators must
sit on the boundary of the worldsheet (see Fig.\ 11). An
exponentiation of the one-loop quark-gluon amplitude would have too
many quark propagators to form a continuous worldsheet where quark
propagators lay on its boundary only.\foot{We thank Howard Schnitzer
for raising that point.}

Assuming our prediction is true, there is another more technical
argument we can make for the non-exponentiation of the planar
one-loop quark amplitude. That is, the six gluon one-loop amplitude
contains a dilogarithm term ($\li2(1+s/t)$ in \mainresult), whereas
the one-loop amplitude with four massless external legs cannot have
dilogarithm. This can be seen by noting that such an amplitude can be
written as a linear combination of scalar one-loop amplitudes with
coefficients that are at most rational functions of the Mandelstam
variables. The scalar one-loop amplitudes that can appear are the
scalar box amplitude with four massless external legs, scalar
triangle amplitude with one massive and two massless external legs,
and the scalar `diangle' amplitude, \ie\ the scalar one-loop amplitude with
two propagators and two massive
external legs. Since none of these scalar one-loop amplitudes gives a
dilogarithm, the one-loop two quarks to two gluons amplitude (times
the tree-level one) cannot match the corresponding six gluon
one-loop amplitude.

Next, we would like to check if our conjectural relation to the
gluon amplitude could hold order by order in perturbation theory.
What could extrapolation to small $\lambda$ of the relation between
the two quarks - two gluon amplitude and the six gluon amplitude
mean? At least naively, it means that order by order in perturbation
theory, the quark-gluon amplitude is the square root of the six
gluon one (see Fig.\ 13). That is obviously wrong as can be seen for
example by noting again that the one-loop quark gluon amplitude does
not contain a dilogarithm, whereas the six gluon one at one and two
loops amplitude does.
\fig{{\bf a)} Diagrammatic form of our strong coupling conjecture.
{\bf b)} One of the diagrams that would contribute in a naive
extrapolation of our conjecture to one-loop.}{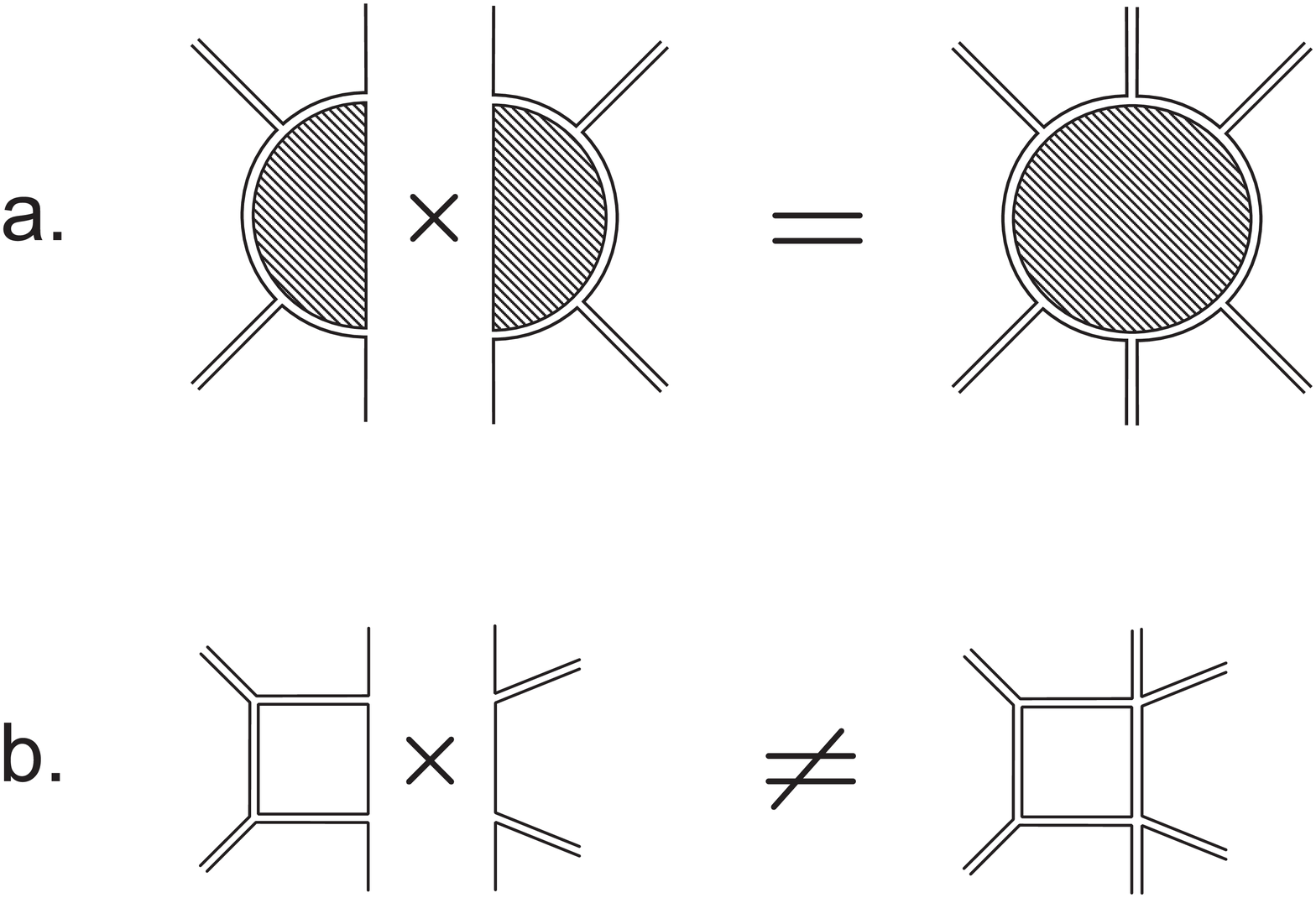}{3in}
It would be nice to have a prediction for the all-orders amplitudes
in the $\NN=2$ theory.

\subsec{Regge behavior of quark scattering amplitudes in $\NN=2$ SYM}

In the previous subsection we have learned that the planar one-loop
massless quark amplitudes in $\NN=2$ SYM (in the probe approximation)
does not exponentiate. We have also seen that naive extrapolation of
our strong coupling conjectural relation to one-loop would
contradict our prediction. A comparison of our strong coupling
prediction with (weak coupling) perturbative results may still be
possible in some special limits, where an all-loop perturbative
computation is in reach. Such a limit is the Regge limit. In this
limit it is believed that, by summing over ladder diagrams
\SchnitzerRN, one can obtain an all-loops planar result.

Therefore, in this section we investigate the Regge limit of the
on-shell planar two quarks~$\to$~two gluons amplitude in strongly
coupled $\NN=2$ SYM, using our explicit prediction \mainresult. The
Regge limit is the limit where the center of mass energy squared is
taken to infinity (and is therefore timelike) with fixed spacelike
momentum transfer.

In the notation of section 5, the Mandelstam variables are
$s=-(\bk_1+\bk_2)^2$, $t=-(\bk_2+\bk_3)^2$ and $u=-(\bk_2+\bk_4)^2$,
where $(\bk_1,\bk_2,\bk_3,\bk_4)$ are the momentum of
$(q_1,g_2,g_3,\bar q_4)$ correspondingly. There are four different
Regge limits we can take here:\foot{The other two possible limits will be
discussed below.}

\bigskip

\item{1)}{$q_1$ and $g_3$ are incoming and $g_2$, $\bar q_4$ are
outgoing. In that case center of mass energy squared is
$u\to\infty$, whereas the fixed spacelike momentum transfer is $s<0$
(therefore $t\to -\infty$). That is the convention in which $u$, $s$
and $t$ are usually used. Here, to keep things short, we will use
the same expressions for $u$, $s$ and $t$ in terms of the momenta in
the three other cases as well.}
\item{2)}{$q_1$ and $g_3$ are incoming and $g_2$, $\bar q_4$ are
outgoing, $u\to\infty$ with fixed $t<0$.}
\item{3)}{$q_1$ and $g_2$ are incoming and $g_3$, $\bar q_4$ are
outgoing, $s\to\infty$ with fixed $t<0$.}
\item{4)}{$q_1$ and $\bar q_4$ are incoming and $g_2$, $g_3$ are
outgoing, $t\to\infty$ with fixed $s<0$.}

\bigskip

An amplitude is said to have Regge behavior if in the Regge limit
it approaches
\eqn\Regge{\AA_{qgg\bar q}(a,b)=\beta(b)\({a\over
-b}\)^{\alpha(b)}\[1+\OO(|b|/a)\]~,}
where $a\in\{u,s,t\}$ is the center of mass energy, $b\in\{u,s,t\}$
is the fixed spacelike momentum transfer, $\alpha(b)$ is the Regge
trajectory and $\beta(b)$ is the Regge residue.

In \DrummondAU\ and \NaculichUB, assuming the BDS ansatz, the planar
four-gluon amplitude in $\NN=4$ SYM was found to possess Regge
behavior, which was further analyzed. In \DrummondAU, the Regge
limit of the $in\to in\to out\to out$ amplitude was analyzed using
\bds. It was found that the amplitude is Regge exact.\foot{That is,
for any values of $t$ and $s$ it can be written in the Regge form
\Regge, with no subleading corrections ($\OO(|t|/s)$ in \Regge).} In
\NaculichUB, based on the BDS ansatz and the strong coupling
prediction \AldayHR, leading Regge trajectory was found together
with an infinite number of daughter trajectories and analyzed in the
$in\to out\to in\to out$ amplitude. Next we will see that the
two-quark -- two-gluon amplitude has Regge behavior in four
different channels (but is never Regge exact).

It follows from \Regge\ that in the Regge limit, the Regge
trajectory is given by
\eqn\Rtraject{\alpha(b)=-{\d\over\d\ln a}
\SS_{\bar qggq}(a,b)
~,}
(recall that $\SS \equiv -\ln \CA$). Note that an amplitude is
dominated by a Regge trajectory in the Regge limit only if the
leading term on the right hand side of \Rtraject\ is
$a$-independent. To analyze our prediction for the amplitude
\mainresult\ in the Regge limit, we need the asymptotic expansion of
the term $\li2\(1+x\)$ in the limits $|x|\to\infty$ and $|x|\to 0$.
For that aim we use the relations
\eqn\relation{\eqalign{&\li2(1+x)+\li2(1+x^{-1})=-\half\ln^2(-x)~,\cr
&\li2(-x)+\li2(1+x)={1\over 6}\pi^2-\ln (-x)\ln(1+x)~,\cr
&\li2(x)=\sum_{k=1}^\infty{x^k\over k^2}~.}}
For $|x|\to\infty$ we combine these as
\eqn\asymptoticst{\li2\(1+x\)=-\half\ln^2\(-x\)+\ln\(-x^{-1}\)\ln\(1+x^{-1}\)-{1\over
6}\pi^2+\sum_{k=1}^\infty{(-x)^{-k}\over k^2}}
whereas, for $|x|\to 0$
\eqn\asymptoticts{\li2\(1+x\)={1\over
6}\pi^2-\ln\(-x\)\ln\(1+x\)-\sum_{k=1}^\infty{\(-x\)^k\over k^2}~.}
By plugging these equations into \mainresult\ we see that in each of
the Regge limits, the $\ln^2(a)$ term exactly cancels. We list below
the results for $\alpha_3(s)$ and $\alpha_4(s)$. The two other Regge
trajectories and the four finite parts can be easily computed,
however we will not need them here. We find that
\eqn\Reggeresult{\eqalign{\alpha_3(t)=&-{1\over
4}f(\lambda)\ln\({t\over\mu^2}\)+\half g(\lambda)+{1\over
4\epsilon}f^{(-1)}(\lambda)+{\rm [dressing~contribution]} \cr
\alpha_4(s)=& -{1\over 8}f(\lambda)\ln\({4s\over\mu^2}\)+{1\over
4}g(\lambda)+{1\over 8\epsilon}f^{(-1)}(\lambda)+{\rm
[dressing~contribution]} ~,}}
where $\lambda{\d\over\d\lambda}f^{(-1)}(\lambda)=f(\lambda)$ and
the [dressing contribution] is the contribution from the
helicity-dependent kinematic factors that multiply the exponential.

First, we claim that the fact that our prediction leads
to Regge trajectories in the Regge limit is a non-trivial
consistency check. The reason is the following.  These amplitudes are IR
divergent and need an IR regulator. One such IR regulator is
implemented by introducing a mass-gap \SchnitzerRN. In the Regge
limit one takes the UV limit where the center of mass energy diverges
(with fixed 4-momentum transfer), keeping the IR regulator fixed.
There is a general proof that renormalizable non-abelian gauge
theories with a mass gap, lead to Regge trajectories for the
elementary fields of the theory \Mandelstam.\foot{Moreover, in such
theories $\alpha(0)$ is the elementary spin (that is, 1 for vectors
and 1/2 for fermions).} In the probe approximation, the $\NN=2$ theory
at hand is conformal and therefore renormalizable. The existence of
these trajectories should be independent of the choice of IR
regulator, although the precise behavior of the trajectory itself
($\alpha(b)$), in the IR region, will depend on the details of the
IR regulator chosen for the massless fields.\foot{We thank Howard
Schnitzer for explaining the related facts to us.}

As stated in the beginning of this section, in the Regge limit we
can actually compare terms in \Reggeresult\ with a perturbative
all-loop computation. These are the coefficients of the leading log
contributions to the Regge trajectories, $-{1\over 4}f(\lambda)$ and
$-{1\over 8}f(\lambda)$ in $\alpha_3$ and $\alpha_4$
correspondingly. In the perturbative region these become
$-{\lambda\over 8\pi^2}$ and $-{\lambda\over 16\pi^2}$. In planar
$\NN=4$ SYM, the leading log coefficient of the Regge trajectory is
$-{1\over 4}f(\lambda)$. In the perturbative regime, it is captured
by a specific infinite subclass of diagrams -- the ladder graphs
\SchnitzerRN. A generic $\NN=4$ ladder diagram that contributes to
the four gluon Regge trajectory in the s-channel is drawn in Fig.\
14.a. For any given ladder diagram, if we replace two of the
external gluons on the left with quark anti-quark pair, as drawn in
Fig.\ 14.b, then only the first `rung' of the ladder is changed and
the rest of the ladder remains the same. After summing the leading
logs of all these ladder diagrams, such a change can only affect the
dressing factor in front of the exponent. We therefore conclude that
in case 3) of the quark-gluon Regge limit considered above, the
leading log contributions to the Regge trajectory must be the same
as the one in the $\NN=4$ gluon Regge trajectory -- in agreement
with what we have found \Reggeresult. If instead of case 3), we
consider case 4), then each of the ladder diagram looks as in Fig.\
14.c. In that case the basic block in the ladder is changed. We
conjecture that these still sum to give the leading log
contributions to the corresponding Regge trajectory but leave the
check for future work. If true, following our results for
$\alpha_4$, we expect it to be one-half the one for gluons
($-{\lambda\over 8\pi^2}$).
\fig{{\bf a)} A generic ladder diagram that contribute to the
leading $log$ in the $\NN=4$ planar Regge trajectory. {\bf b, c)}
Generic ladder diagrams in quark gluon planad
amplitude.}{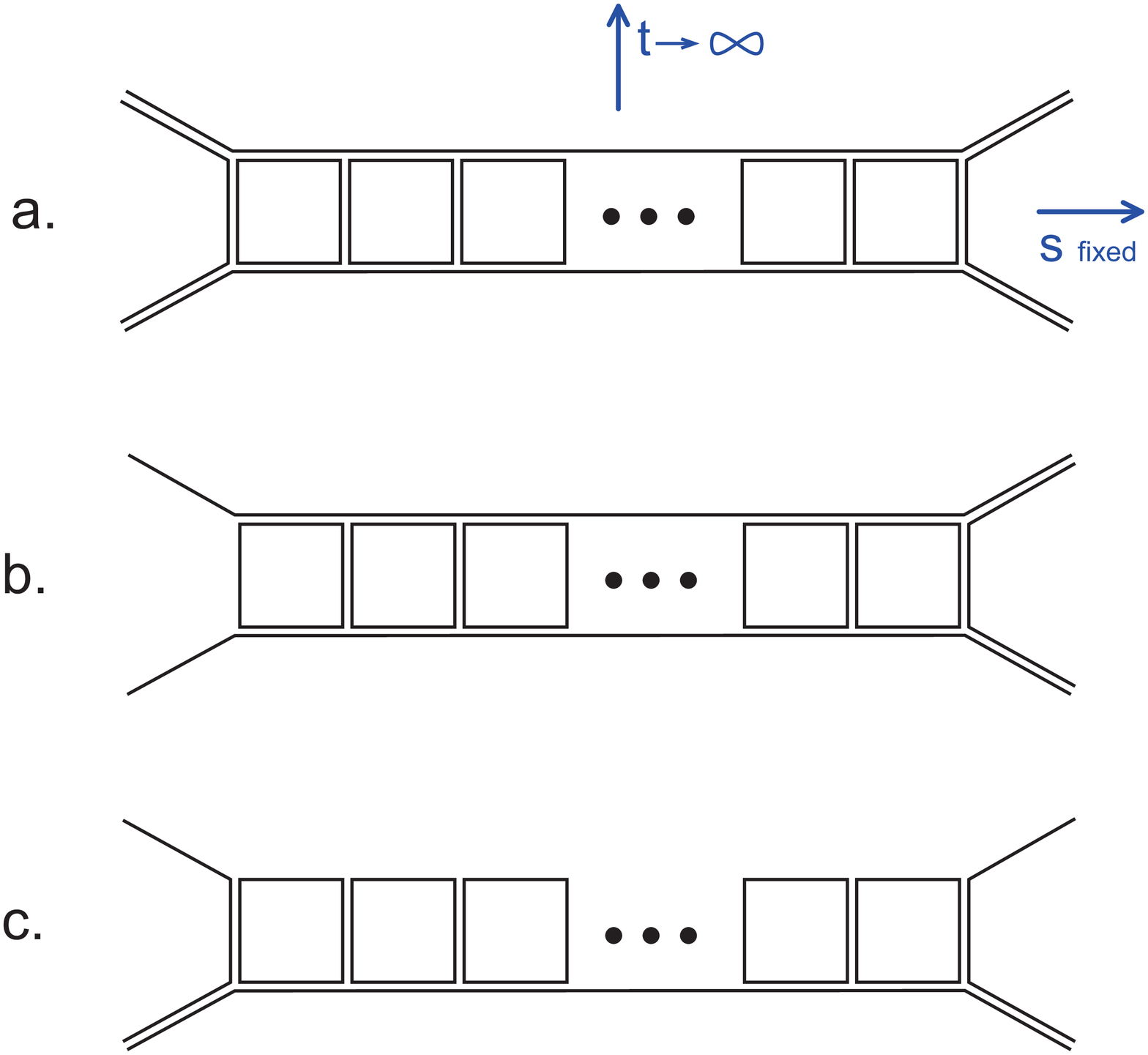}{3in}

There are other two Regge limits we can take that are not listed
above. These are the limits where $s\to\pm\infty$ with $u<0$ fixed.
In these limits the fixed momentum transfer is not between two
color-adjacent partons. Therefore, the ladder diagrams that
perturbatively lead to the Regge trajectory are not planar. We then
expect the coefficent in front of the corresponding Regge term (that
dominates the amplitude at finite $N$, $|s|\to \infty$) to scale as
$1/N$ and to be absent in the strict planar limit. That is, the
amplitude takes the rough form
\eqn\planarsuppression{\CA_{{\rm other~two~channels}} \sim
{\gamma(b)\over N}\({a\over -b}\)^{\alpha(b)} + \eta(b)e^{ -
\chi(\lambda) \ln^2\(-a/b\)}~,}
where the functions $\gamma(b)$ and $\eta(b)$ scales as $N^0$ in the
large $N$ limit. Indeed, only in these two cases, our prediction
\mainresult\ does not lead to Regge behavior in the Regge limit.

\newsec{Brief conclusions}

In this paper, we generalized the prescription of Alday and Maldacena
\AldayHR\foot{Other related papers include
\otherrefs.} to
include fields in the fundamental representation.
A direct solution of the resulting extremal-area problem
is hard.  We circumvented this difficulty
by relating its solution to an auxiliary $\NN=4$
amplitude known from \bds.

The resulting answer passed several checks. The IR divergent parts
reproduce the Sudakov factors of a large-$N$ theory with
fundamentals. The answer has planar Regge behavior in the channels
where it should. In one channel (See Fig.\ 14.b), the leading log of
the trajectory matches that of the $\NN=4$ theory as it should.

Our main result, then, is a relationship between amplitudes
in two field theories.
It would be nice to have some direct understanding of the
origin of this relation.

\bigskip

{\bf Note added}

After this paper was published, \AldayHE\ suggested that the BDS
ansatz should be corrected for large numbers of gluons. Although
some of our consistency checks assume the BDS ansatz for the six
gluons amplitude, the main point of our paper, relating scattering
amplitudes with quarks to pure gluon amplitudes, is independent of
that ansatz.

\bigskip
\centerline{\bf{Acknowledgements}}

We thank A. Lawrence, H. Liu, J. Maldacena, H. Schnitzer for
discussions and comments, and H. Elvang and D. Freedman for help
with the literature. A.S. would like to thank the MIT Center for
Theoretical Physics for their generous hospitality. The work of J.M.
is supported in part by funds provided by the U.S. Department of
Energy (D.O.E.) under cooperative research agreement
DE-FG0205ER41360. The work of A.S. is supported in part by the NSF
grant PHY-0331516, by DOE Grant No. DE-FG02-92ER40706, and by an
Outstanding Junior Investigator award.

\appendix{A}{The one-loop finite remainder, $F^{(1)}_n$}

The one-loop finite remainder, $F^{(1)}_n$ appearing in \finite, was
evaluated in \BernZX. It is given by (at $\epsilon=0$):
\eqn\Fone{F^{(1)}_n=\half\sum_{i=1}^n\[D_{n,i}
+L_{n,i}-\sum_{r=2}^{[n/2] -1} \ln \({\tn{r}{i}\over \tn{r+1}{i} }\)
\ln \({\tn{r}{i+1}\over\tn{r+1}{i} }\) + {3\over2} \zeta_2\]}
where $[x]$ is the greatest integer less than or equal to $x$. Here
$\tn{r}{i} = -(k_i + \cdots + k_{i+r-1})^2$ are the momentum
invariants, so that $\tn1i = 0$ and $\tn2i = s_{i,i+1}$. (All
indices are understood to be $\mod\ n$.) The form of $D_{n,i}$ and
$L_{n,i}$ depends upon whether $n$ is odd or even. For $n=2m+1$,
\eqn\DLodd{\eqalign{D_{2m+1,i} &=-\sum_{r=2}^{m-1} \li2 \(1-{
\tn{r}{i} \tn{r+2}{i-1} \over \tn{r+1}{i} \tn{r+1}{i-1}}\)~,\cr
L_{2m+1,i} &=-{ 1\over 2}
  \ln\({\tn{m}{i}\over\tn{m}{i+m+1}}\)\ln\({\tn{m}{i+1}\over\tn{m}{i+m}}\)~,}}
whereas for $n=2m$,
\eqn\DLeven{\eqalign{D_{2m,i} &=-\sum_{r=2}^{m-2} \li2 \(1-{
\tn{r}{i} \tn{r+2}{i-1} \over \tn{r+1}{i} \tn{r+1}{i-1}}\) -{1 \over
2} \li2 \(1-{\tn{m-1}{i} \tn{m+1}{i-1} \over \tn{m}{i}
\tn{m}{i-1}}\)~,\cr L_{2m,i} &=-{1\over 4}\ln\({\tn{m}{i}\over
\tn{m}{i+m+1}}\)\ln\({\tn{m}{i+1}\over\tn{m}{i+m}}\)~.}}

\bigskip
\listrefs
\end